%% file: main.tex
\documentclass[10pt,conference]{style/IEEEtran}

\input{tex/env}

\title{Duet: Creating Harmony between Processors \\ and Embedded FPGAs} 

\author{
    \IEEEauthorblockN{Ang Li}
    \IEEEauthorblockA{
    Princeton University \\
    Princeton, NJ 08544, USA \\
    angl@princeton.edu
    }
    \and
    \IEEEauthorblockN{August Ning}
    \IEEEauthorblockA{
    Princeton University \\
    Princeton, NJ 08544, USA \\
    aning@princeton.edu
    }
    \and
    \IEEEauthorblockN{David Wentzlaff}
    \IEEEauthorblockA{
    Princeton University \\
    Princeton, NJ 08544, USA \\
    wentzlaf@princeton.edu
    }
}

\begin{document}
\maketitle
\thispagestyle{plain}
\pagestyle{plain}

\input{tex/abstract}
\input{tex/introduction}
\input{tex/architecture}
\input{tex/application}
\input{tex/dolly}
\input{tex/evaluation}
\input{tex/related}
\input{tex/conclusion}

\section*{Acknowledgements}

This material is based on research sponsored by the Air Force Research Laboratory (AFRL) and Defense Advanced Research Projects Agency (DARPA) under agreement No. FA8650-18-2-7852. This material is based upon work supported by the National Science Foundation under Grant No. CNS-1823222 and the National Science Foundation Graduate Research Fellowship Program under Grant No. DGE-2039656. The U.S. Government is authorized to reproduce and distribute reprints for Governmental purposes notwithstanding any copyright notation thereon. The views and conclusions contained herein are those of the authors and should not be interpreted as necessarily representing the official policies or endorsements, either expressed or implied, of Air Force Research Laboratory (AFRL) and Defense Advanced Research Projects Agency (DARPA) or the U.S. Government.  Any opinions, findings, and conclusions or recommendations expressed in this material are those of the author(s) and do not necessarily reflect the views of the National Science Foundation.

\balance
\newpage

\bibliographystyle{IEEEtranS}
\bibliography{refs}

\end{document}

%% file: tex/env.tex
\usepackage{cite}
\usepackage{amsmath,amssymb,amsfonts}
\usepackage{algorithmic}
\usepackage{graphicx}
\usepackage{textcomp}
\usepackage{xcolor}
\usepackage[hyphens]{url}

\def\BibTeX{{\rm B\kern-.05em{\sc i\kern-.025em b}\kern-.08em
    T\kern-.1667em\lower.7ex\hbox{E}\kern-.125emX}}

\usepackage{soul}
\usepackage{multirow}
\usepackage{multicol}
\usepackage{enumitem}
\usepackage{listings}
\usepackage[T1]{fontenc}
\usepackage{inconsolata}
\usepackage[utf8]{inputenc}
\usepackage{threeparttable}
\usepackage{pifont}
\usepackage{balance}
\usepackage{ragged2e}
\usepackage{subcaption}
\usepackage[export]{adjustbox}
\usepackage{tikz}
\usetikzlibrary{shapes.geometric}
\usepackage{lipsum}

\paperwidth=\pdfpagewidth
\paperheight=\pdfpageheight

\usepackage{hyperref}
\hypersetup{bookmarks=true,breaklinks=true,colorlinks,citecolor=blue,linkcolor=blue,urlcolor=blue}

\newcommand{\circlenum}[1]{\raisebox{.5pt}{\textcircled{\raisebox{-.9pt} {#1}}}}

\newif\ifrebuttal

\ifrebuttal
\newcommand{\rebut}[1]{\hl{#1}}

\else 
\newcommand{\rebut}[1]{#1}

\fi

%% file: tex/abstract.tex
\begin{abstract}
    
The demise of Moore's Law has led to the rise of hardware acceleration.
However, the focus on accelerating stable algorithms in their entirety neglects the abundant fine-grained acceleration opportunities available in broader domains and squanders host processors' compute power.

This paper presents Duet, a scalable, manycore-FPGA architecture that
promotes embedded FPGAs (eFPGA) to be equal peers with processors through non-intrusive, bi-directionally cache-coherent integration.
In contrast to existing CPU-FPGA hybrid systems in which the processors play a supportive role, Duet unleashes the full potential of both the processors and the eFPGAs with two classes of post-fabrication enhancements:
\textit{fine-grained acceleration}, which partitions an application into small tasks and offloads the frequently-invoked, compute-intensive ones onto various small accelerators, leveraging the processors to handle dynamic control flow and less accelerable tasks;
\textit{hardware augmentation}, which employs eFPGA-emulated hardware widgets to improve processor efficiency or mitigate software overheads in certain execution models.

An RTL-level implementation of Duet is developed to evaluate the architecture with high fidelity.
Experiments using synthetic benchmarks show that Duet can reduce the processor-accelerator communication latency by up to 82\% and increase the bandwidth by up to 9.5x.
The RTL implementation is further evaluated with seven application benchmarks, achieving 1.5-24.9x speedup.
    
\end{abstract}

%% file: tex/introduction.tex
\section{Introduction} \label{sec:introduction}


\input{figures/arch_comparison}

The demise of Moore's Law~\cite{end-of-moore} and the stagnation of processor performance growth~\cite{cpu-scaling} have given rise to hardware acceleration~\cite{dl-asic, eyeriss, asic-cloud}.
Since ASIC-based accelerators suffer from high non-recurring engineering costs and low versatility, field-programmable gate arrays (FPGAs) are gaining steam due to their post-fabrication, gate-level reconfigurability and close-to-ASIC performance~\cite{fpga-vs-cpu-berkley, fpga-vs-cpu-cmu}, \cite{fpga-asic-gap-dl}.
FPGAs can be integrated either as standalone devices (Fig.~\ref{fig:arch-comp-off-chip}) or into field-programmable system-on-chips (FPSoC) (Fig.~\ref{fig:arch-comp-fpsoc}).
The conventional, FPGA-based acceleration paradigm is \textbf{coarse-grained acceleration} (Fig.~\ref{fig:hardware-acceleration}b), which offloads an algorithm in its entirety onto the FPGA.
Despite offering substantial improvements in performance and energy efficiency, coarse-grained acceleration is only practically applicable to limited algorithms that are both stable (to justify the accelerator design costs) and sufficiently large (to justify the control and data transfer overheads).

\input{figures/hardware_acceleration}

This work proposes \textbf{Duet} (Fig.~\ref{fig:arch-comp-duet}, Sec.~\ref{sec:architecture}), a novel, cache-coherent, manycore-FPGA architecture that is tailored to enable two novel paradigms of hardware acceleration.
\textbf{\mbox{Fine-grained} \mbox{acceleration}} (Fig.~\ref{fig:hardware-acceleration}c, Sec.~\ref{sec:fine-grained-acc}) partitions an algorithm into smaller tasks and offloads only the frequently-invoked, compute-intensive ones onto a variety of small accelerators.
Processors still play a critical role by handling dynamic control flow, memory/IO-bound tasks, or any other less accelerable computations.
For example, fine-grained accelerators can be used for special instructions like tangent, basic algorithms like sorting, or inner loop bodies like the compute payload per node/edge during a graph traversal.
\textbf{Hardware augmentation} (Fig.~\ref{fig:hardware-acceleration}d, Sec.~\ref{sec:hardware-aug}) takes an application-agnostic approach \textemdash~it employs FPGA-emulated hardware widgets to reduce processor idle time or mitigate software overhead in certain execution models.
For example, hardware-implemented, lock-free data structures can reduce synchronization overhead in shared-memory, multi-thread programs;
hardware task schedulers enable task-level parallelism with lower overhead than software schedulers.

Existing CPU-FPGA systems are inefficient for fine-grained acceleration and hardware augmentation due to two reasons:
first, the centralized, bandwidth-optimized CPU-FPGA interconnect performs poorly when lots of cores and accelerators communicate via short, frequent messages;
second, even on state-of-the-art FPSoCs which support bi-directional cache coherence between the processors and the FPGA, cacheline-granular memory sharing incurs non-trivial overhead.
The overhead is justifiable for coarse-grained acceleration due to extremely high compute-to-memory ratio, but becomes critical for small, frequently-invoked, fine-grained accelerators.

\textbf{Duet} is tailored to the distinct architecture requirements posed by fine-grained acceleration and hardware augmentation.
In essence, \textbf{Duet promotes eFPGAs to be equal peers with processors and integrates them as first-class citizens on the network-on-chip (NoC) through the novel, lightweight, Duet Adapters}.
In particular, Duet contains the following novelties:

\begin{itemize}[leftmargin=0em, itemindent=2em]
    \item \textbf{Scalability}:
    Duet enables tight integration of one to multiple eFPGAs of various resource compositions into a scalable manycore system.
    Each eFPGA may be connected to multiple Duet Adapters for higher aggregate memory bandwidth.

    \item \textbf{Hybrid Cache Coherence}:
    Duet adopts a hybrid scheme to coherently integrate the eFPGAs.
    Each Duet Adapter contains one to multiple private, local, hardware \textbf{Proxy Caches} (Sec.~\ref{sec:arch-proxy-cache}) that translate the platform-dependent, cache coherence protocols into simple memory interfaces for the eFPGA.
    Furthermore, each Proxy Cache can be configured at eFPGA programming time to support an optional, bi-directionally coherent, soft cache built out of eFPGA resources.
    The use of soft caches improves accelerator performance by exploiting data locality while coherently sharing data with the processors.
    
    \item \textbf{Non-Intrusive Integration}:
    For fine-grained acceleration and hardware augmentation, a sizable amount of compute is still run on the processors and is critical to the overall performance.
    However, the direct participation of eFPGA-emulated, soft caches in cache coherence may slow down the cache system because eFPGAs run at a lower clock frequency and suffer from clock-domain-crossing overheads.
    Addressing this challenge, the Duet Adapters operate in the processors' (fast) clock domain and move FPGA-side coherence maintenance off of the critical path.
    As a result, the cache system performs equally fast as if each Duet Adapter was an additional processor-owned private cache.


    \item \textbf{Plug-and-Play Integration}:
    Duet requires little to no hardware changes to existing manycore systems because the Duet Adapters transduce between the eFPGAs and the NoC.
    Such decoupling is critical because modifying a mature processor design often leads to performance degradation or even hardware bugs, especially as design complexity and verification costs skyrocket on advanced technology nodes~\cite{vlsi-cost}.

\end{itemize}

To evaluate Duet with high fidelity, we build Dolly,
a prototype instance of Duet at the RTL level (Verilog \& SystemVerilog).
In addition, we assemble a full toolchain for software development and accelerator design, leveraging an array of open-source projects, including OpenPiton~\cite{openpiton}, BYOC~\cite{byoc}, 
PRGA~\cite{prga}, Yosys~\cite{Yosys}, and VTR~\cite{vtr}.
\textbf{To encourage further research in tightly-integrated, hardware cache-coherent CPU-eFPGA systems, we have open-sourced Dolly and its toolchain, available at \url{https://github.com/PrincetonUniversity/Duet}}.

Experiments show that the Duet Adapter introduces negligible hardware overhead for the CPU-eFPGA integration (Sec.~\ref{sec:eval-arch}).
The Proxy Cache reduces the latency of processor-accelerator communication up to 82\% and increases the bandwidth up to 9.5x compared to having the eFPGA participating directly in the coherence protocol 
(Sec.~\ref{sec:eval-itx}).
The latency reduction and bandwidth increase are stable across different eFPGA clock frequencies.
On selected benchmarks (Sec.~\ref{sec:eval-app}), Dolly improves the overall performance up to 24.9x in comparison to corresponding processor-only baselines and up to 4x in comparison to other CPU-FPGA systems.

\input{figures/duet_architecture}

The major contributions of this work are:

\begin{itemize}[leftmargin=0em, itemindent=2em]
    \item
    Presenting Duet, a manycore-FPGA architecture which employs novel Duet Adapters to integrate embedded FPGAs in a scalable, non-intrusive, cache-coherent manner.
    \item
    Identifying and demonstrating with examples two novel paradigms of hardware acceleration enabled by Duet, namely fine-grained acceleration and hardware augmentation. 
    \item
    Presenting Dolly, an RTL-level prototype of Duet and a full toolchain for software development and accelerator design.
    \item
    Evaluating Dolly's silicon area consumption and performance with synthetic and application benchmarks.
    \item
    Releasing Dolly and its toolchain for open-source access.
\end{itemize}

%% file: figures/arch_comparison.tex
\begin{figure}[!t]\centering
    
    \begin{subfigure}[t]{\linewidth}
        \centering
        \includegraphics[width=0.84\linewidth]{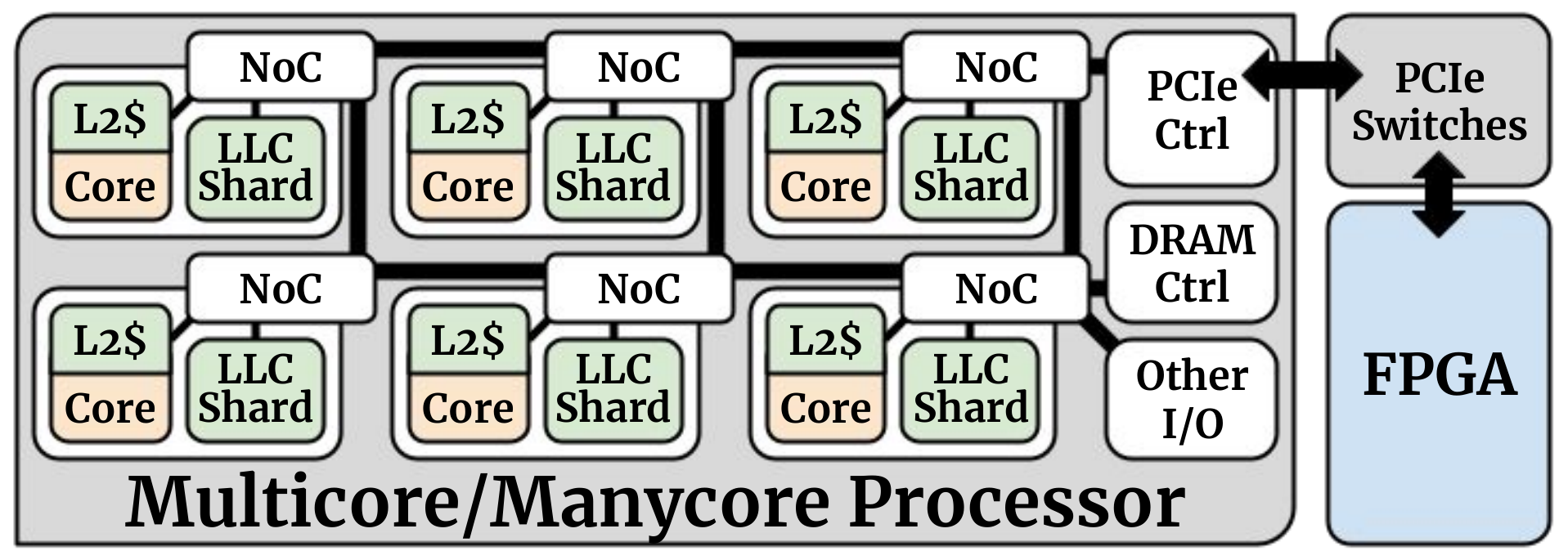}
        \setlength{\abovecaptionskip}{2pt}
        \caption{
        \fontsize{9}{9}\selectfont
        Multicore/Manycore Processor and Standalone FPGA}
        \begin{FlushLeft}
        \fontsize{9}{9}\selectfont
        \vspace{-0.6\baselineskip}
        \textbf{L2\$}=Private cache;
        \textbf{NoC}=Network-on-chip;
        \textbf{LLC}=Last-level cache
        \end{FlushLeft}
        \label{fig:arch-comp-off-chip}
    \end{subfigure}

    \begin{subfigure}[t]{\linewidth}
        \centering
        \includegraphics[width=0.84\linewidth]{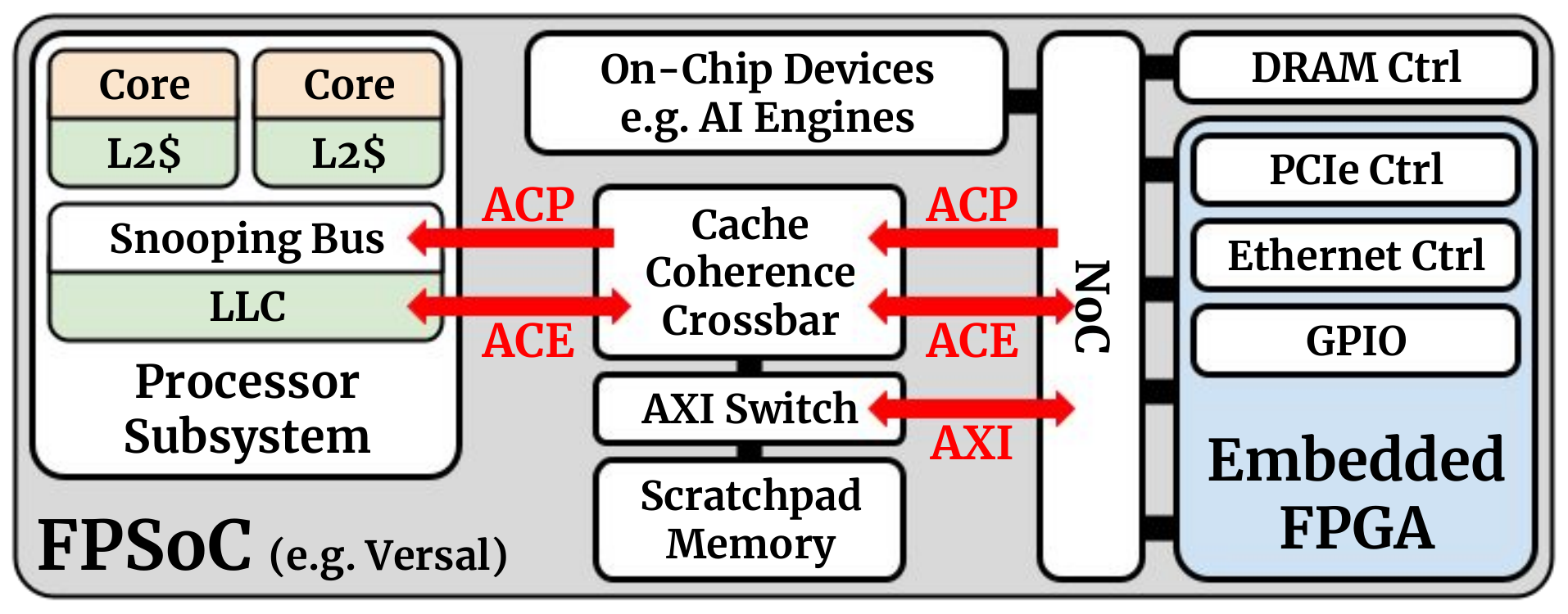}
        \setlength{\abovecaptionskip}{2pt}
        \caption{
        \fontsize{9}{9}\selectfont
        Field-Programmable System-on-Chip (FPSoC), e.g. Versal~\cite{xilinx-acap}
        }
        \begin{FlushLeft}
        \fontsize{9}{9}\selectfont
        \vspace{-0.6\baselineskip}
        AXI~\cite{axi4}: non-coherent interconnect;
        ACP~\cite{axi4}: uni-directional coherent interconnect;
        ACE~\cite{axi4}: fully coherent interconnect.
        \end{FlushLeft}
        \label{fig:arch-comp-fpsoc}
    \end{subfigure}

    \begin{subfigure}[t]{\linewidth}
        \centering
        \includegraphics[width=0.84\linewidth]{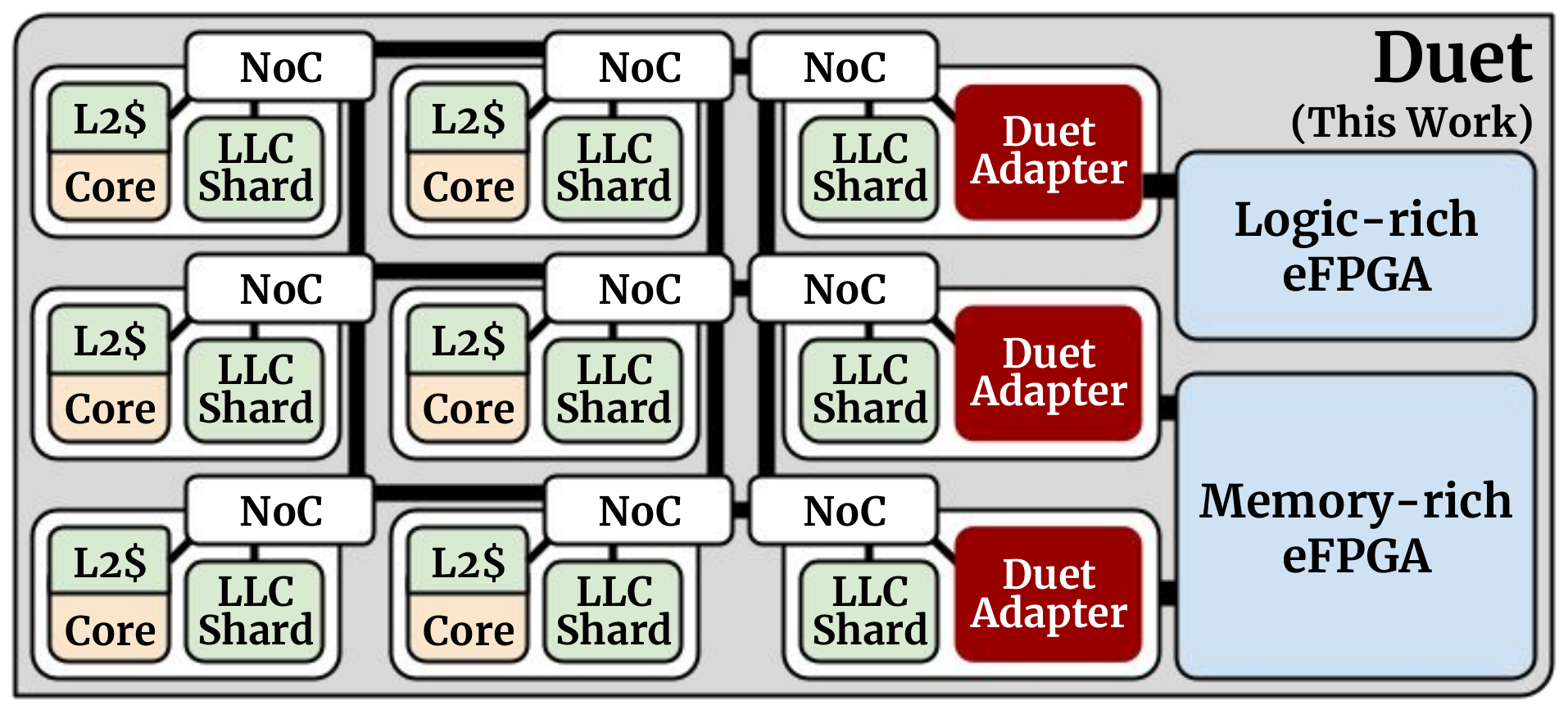}
        \setlength{\abovecaptionskip}{2pt}
        \caption{
        \fontsize{9}{9}\selectfont
        Duet (This Work)
        }
        \begin{FlushLeft}
        \fontsize{9}{9}\selectfont
        \vspace{-0.6\baselineskip}
        Each Duet Adapter can be configured independently to support uni- or bi-directional coherence in addition to a memory-mapped non-coherent interface.
        Duet enables scalable integration of: 1) any number of processors; 2) multiple independent embedded FPGAs (eFPGAs); 3) multiple NoC access points per eFPGA.
        \end{FlushLeft}
        \label{fig:arch-comp-duet}
    \end{subfigure}
    
    \setlength{\belowcaptionskip}{-8pt}
    \setlength{\abovecaptionskip}{-1pt}
    \caption{CPU-FPGA Systems}
    \label{fig:arch-comp}
\end{figure}

%% file: figures/hardware_acceleration.tex
\begin{figure}[t!]
    \centering
    \includegraphics[width=0.95\linewidth, trim={0pt 4pt 0pt 2pt}, clip]{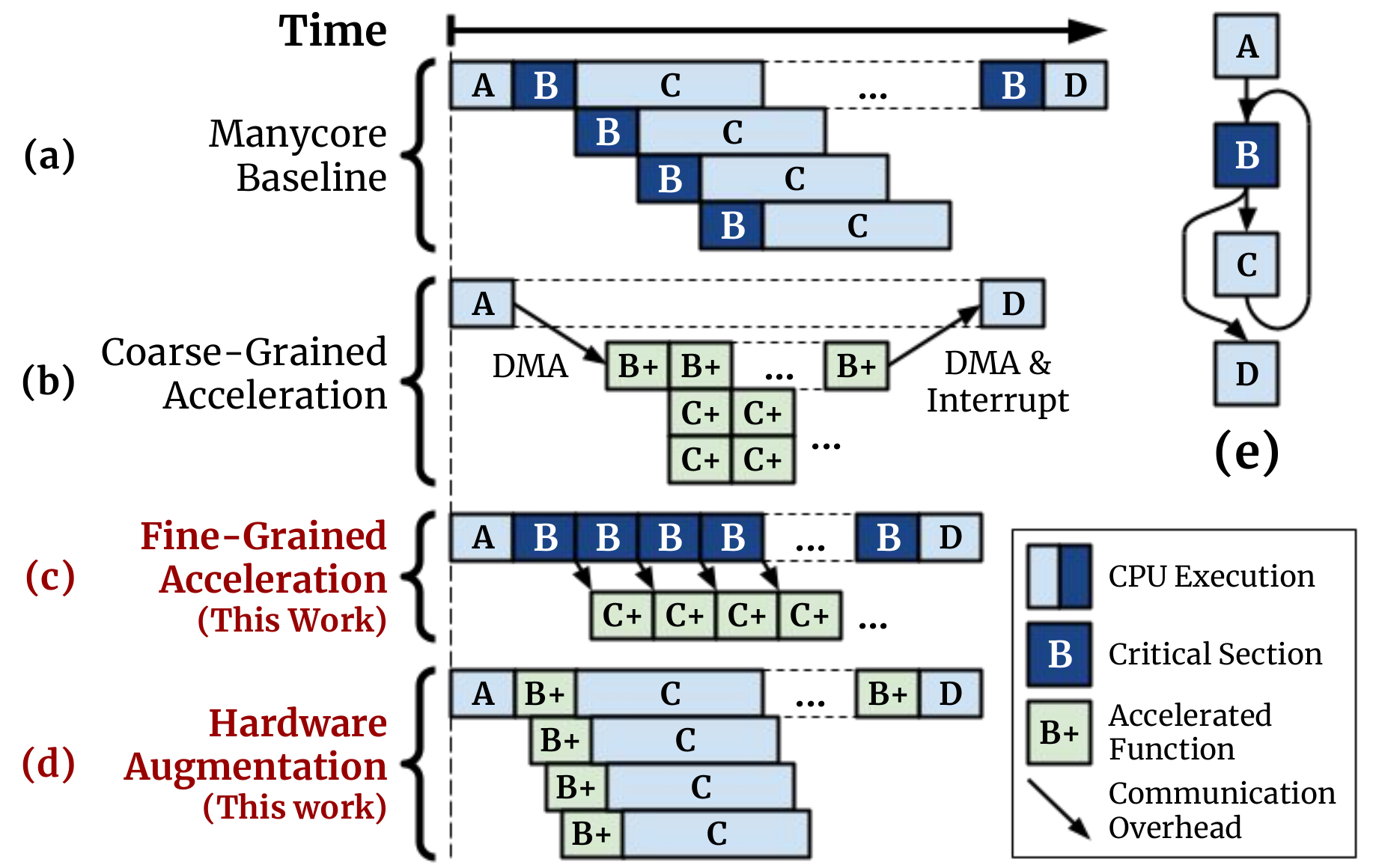}
    \setlength{\abovecaptionskip}{4pt}
    \caption{Accelerating a Hypothetical Program}
    
    \begin{FlushLeft}
    \fontsize{9}{9}\selectfont
    \vspace{-0.5\baselineskip}
    (a-c) Execution time of a manycore baseline and different acceleration paradigms;
    (d) An example of hardware augmentation in which the embedded FPGA emulates a lock-free task scheduler;
    (e) Control flow graph of the program.
    \vspace{-\baselineskip}
    \end{FlushLeft}
    \label{fig:hardware-acceleration}
\end{figure}

%% file: figures/duet_architecture.tex
\begin{figure*}[t!]
    \centering
    \includegraphics[scale=0.4, trim=0pt 3pt 0pt 3pt, clip]{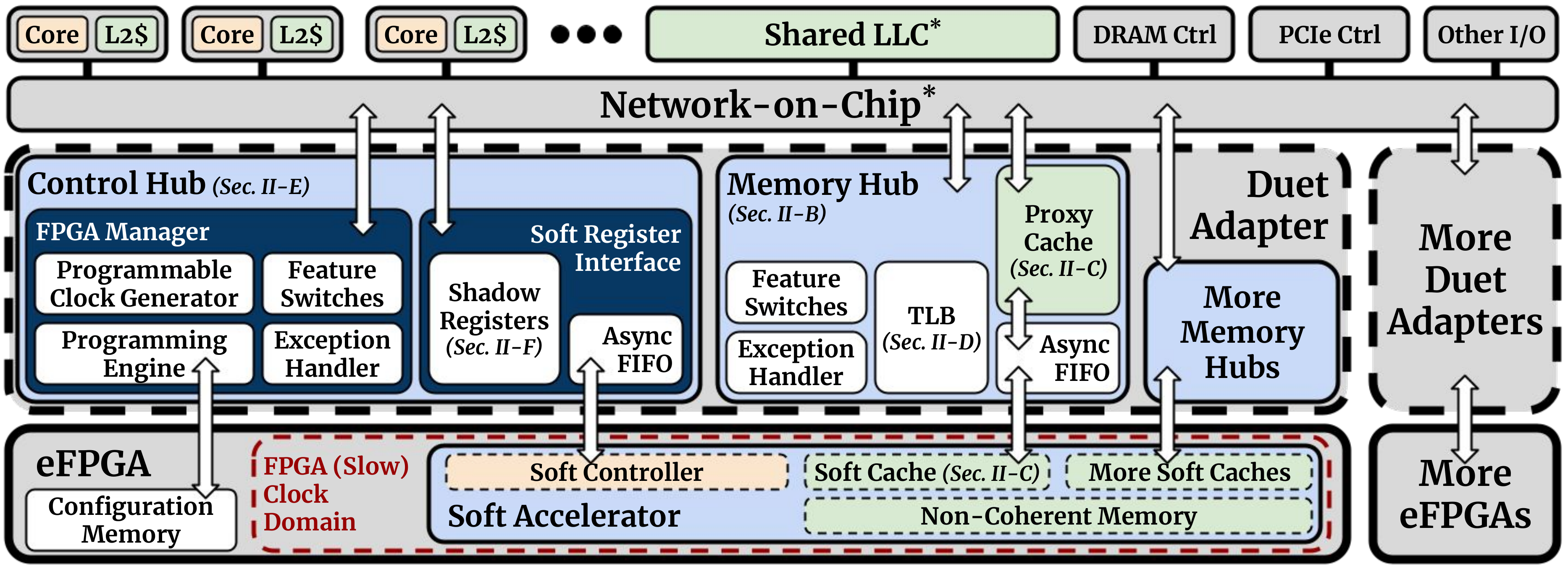}
    \setlength{\abovecaptionskip}{8pt}
    \caption{Duet Architecture and an Emulated Soft Accelerator
    }
    
    \begin{Center}
    \fontsize{9}{9}\selectfont
    \vspace{-0.5\baselineskip}
    * For simplicity, the figure shows a bus-based NoC and a centralized LLC. Duet can be adapted to other NoC topology and distributed LLC.
    For example, Dolly (Sec.~\ref{sec:instance}) uses a 2D-mesh and a distributed LLC.
    \vspace{-0.5\baselineskip}
    \end{Center}
    
    \label{fig:system-architecture}
\end{figure*}

%% file: tex/architecture.tex
\section{Architecture} \label{sec:architecture}

Before diving into the architecture of Duet, we want to clarify several terms that are used throughout this paper:
\textbf{software} refers to the program running on the processors;
\textbf{hardware}, \textbf{hard} cache, etc. refer to the components that are fixed at fabrication time, in contrast to the \textbf{soft} components that are emulated with the reconfigurable resources in the eFPGAs.
In addition, we use the term \textbf{soft accelerator} to refer to both fine-grained accelerators and hardware augmentation widgets.

\subsection{Overview} \label{sec:arch-overview}

Fig.~\ref{fig:system-architecture} shows an overview of the Duet architecture.
Duet
integrates eFPGAs as first-class citizens on the NoC through novel, hardware \textbf{Duet Adapters}.
Each Duet Adapter contains one to several \textbf{Memory Hubs} and one \textbf{Control Hub}.
The Memory Hubs enable bi-directionally coherent memory accesses of the soft accelerators and transduce between the memory interfaces of the eFPGAs and the NoC.
The Control Hubs present the eFPGAs as on-chip devices that are accessible via memory-mapped I/Os (MMIOs).

The fact that eFPGA-emulated soft accelerators typically run at a much slower clock than the rest of the system poses a challenge to minimizing processor-accelerator communication overhead.
In particular, any traffic entering or leaving the slow clock domain must pay for the clock-domain-crossing (CDC) overhead, and every slow clock cycle significantly penalizes the total communication time (Fig.~\ref{fig:cache-soft-only} \& Fig.~\ref{fig:sri-timing-normal}).
A few dozen cycles seem negligible, but since the accelerated tasks are short and invoked frequently, the accumulated overhead can chip away a large fraction of the effective speedup.
Addressing this challenge, the Duet Adapter adopts various strategies to move the logic in the slow clock domain off of the critical path.

\subsection{Memory Hub} \label{sec:arch-memory-tile}

Each Duet Adapter may contain multiple Memory Hubs, each attached to the NoC using an independent connection.
The soft accelerator may use all or any subset of them to access the memory.
Each Memory Hub consists of an exception handler, a set of feature switches, and a \textbf{Proxy Cache} (Sec.~\ref{sec:arch-proxy-cache}), all implemented in hardware.
Besides the hardware Proxy Cache, each memory hub can support one optional, bi-directionally coherent, \textbf{Soft Cache} built out of eFPGA resources.
The exception handler as well as all the feature switches can be configured by the processors via on-chip MMIOs.

The exception handler employs timeout and parity checks to monitor eFPGA outputs.
When an exception is detected, e.g., due to an RTL or software bug, it asserts an error code and deactivates all Memory Hubs in the same Duet Adapter.
Once deactivated, the Memory Hubs stop accepting any memory requests from the eFPGA, but the Proxy Caches remain functional, so that any in-flight coherence messages are processed properly.
This mechanism prevents accelerator bugs from halting the system at the micro-architecture level.

The feature switches allow the processors to configure the Memory Hubs according to the state of the eFPGA and the specifications of the soft accelerator.
For example, the Memory Hubs should be deactivated during eFPGA reconfiguration;
if soft caches are used, the Memory Hubs should be configured to forward invalidation requests into the eFPGA.

\subsection{Proxy Cache} \label{sec:arch-proxy-cache}

\input{figures/cache_options}
\input{figures/l2_timing}

A key challenge in building coherently-integrated CPU-FPGA systems is how to organize the FPGA-side caches.
Fig.~\ref{fig:cache-options} shows the options:

\begin{itemize}[leftmargin=5em]
    \item[\textbf{Soft-Only}]
    The eFPGA is directly connected to the NoC.
    The soft cache, if used, must implement the system-wide cache coherence protocol.
    
    \item[\textbf{Hard-Only}]
    A hardware, private cache is added between the eFPGA and the NoC.
    Cache coherence is hidden from the eFPGA, so soft caches are not supported.
    
    \item[\textbf{Hybrid}]
    A hardware, private cache participates in the system-wide cache coherence on behalf of the eFPGA and supports the use of soft caches with a local cache coherence protocol.
\end{itemize}

Option 1, \textit{soft-only}, has three main drawbacks.
First, it becomes the accelerator developers' burden to design a cache in compliance with a platform-dependent, complex cache coherence protocol.
The sophisticated control logic consumes more logic resources, has higher access latency, and is rarely reusable across devices \textemdash~a soft cache designed for ACE~\cite{axi4} would hardly work with CHI~\cite{chi}.
Second, since the soft caches have access to the micro-architectural state (e.g., can block a NoC message indefinitely), the system cannot fully contain faulty or malicious behaviors of the soft caches.
Third, as shown in Fig.~\ref{fig:cache-soft-only}, the soft caches may slow down the hardware cache system due to CDC overheads and the slow cycles spent in the FPGA clock domain.
Nonetheless, \textit{soft-only} is the de facto design on most commodity FPSoCs because these FPGA-centric architectures are designed for coarse-grained acceleration.
Specifically, coherent memory sharing is rare and mainly offered to simplify programming; soft cache IPs are licensable from the vendors; and accelerator bugs are expected to fail the entire system.

Option 2, \textit{hard-only}, addresses all the drawbacks of the soft-only approach but has other limitations.
First,
the CDC overhead is now imposed on the accelerator datapaths (\ref{fig:cache-hard-only}).
Second, the cache implementation is fixed at fabrication time and may be suboptimal for certain accelerators.

Given the downsides of the soft-only and hard-only options, Duet takes a \textit{hybrid} approach.
A private, local, hardware \textbf{Proxy Cache} implements the platform-dependent, system-wide cache coherence protocol and provides a simple memory interface to the eFPGA.
Each Proxy Cache can be configured through feature switches to support an eFPGA-emulated, soft cache which can be tightly integrated into the accelerator datapaths.
Moreover, the Proxy Cache contains two key novelties when compared to a na\"ive implementation of the hybrid approach:

First, \textbf{the Proxy Cache neither requires nor accepts any acknowledgements from the soft cache}.
As a result, the Proxy Cache always responds to coherence messages promptly (Fig.~\ref{fig:cache-duet}), insulating the cache system from the slow eFPGA clock.
To maintain coherence, the Proxy Cache requires the soft cache to be write-through but allows write buffering (the Proxy Cache itself can be write-back or write-through, depending on the coherence protocol of the LLC).
Furthermore, because the asynchronous FIFOs deliver messages in order, the soft cache always receives invalidations, line fills, and write acks in the same order as they are sent by the Proxy Cache.
Rigid proof of the correctness of such protocols is out of the scope of this paper, but the PCX protocol in OpenSPARC T1~\cite{opensparc} and its extension, the TRI protocol in BYOC~\cite{byoc}, are two examples that meet the above requirements.

Second, the protocol is simple yet flexible.
In the common case, the soft cache only needs to support two request types (\textbf{\texttt{Load}} and \textbf{\texttt{Store}}) and three response types (\textbf{\texttt{LoadAck}}, \textbf{\texttt{StoreAck}} and \textbf{\texttt{Inv}}alidation).
It is up to the accelerator designer whether to use a write buffer, how many entries the write buffer has, and whether read-after-write forwarding is compatible with the consistency assumptions of the application.
The Proxy Cache can be configured to work with either a write-allocate or a write-no-allocate soft cache, as well as be configured to enable atomic operations which require the soft cache to support incrementally more message types.

\subsection{Memory Protection and Virtualization} \label{sec:arch-virt}

Another key challenge for CPU-FPGA systems is memory protection and virtualization.
The Proxy Cache is physically-indexed, physically-tagged because it is implemented in hardware and is closer to the LLC.
However, it depends on the type of soft accelerator whether the eFPGA should use virtual or physical addresses. 
Hardware augmentation widgets may be trusted firmware and can be granted access to physical memory.
On the contrary, application-specific, fine-grained accelerators are like user programs and can be faulty or malicious, so they are better restricted to virtual addresses.

To enable virtual memory accesses of the soft accelerator, Duet adds a translation look-aside buffer (TLB) to each Memory Hub.
The TLB can be disabled if the soft accelerator is granted access to the physical memory space;
otherwise, accelerator-initiated memory accesses must be translated by the TLB while being speculatively processed by the Proxy Cache.
On a page fault, the TLB sends an interrupt to a processor, then the kernel-level interrupt handler either updates the TLB using MMIOs or kills the accelerator if the page access is deemed invalid.

One special case is when soft caches are used but the accelerator only has access to virtual addresses, i.e., the soft caches are virtually-indexed, virtually-tagged.
To enable reverse-mapping from physical address to virtual address when the Proxy Cache forwards an invalidation into the soft cache, the Proxy Cache stores the virtual page number beside the physical tag of each cacheline.
This also rules out the coexistence of synonym aliases (different virtual addresses mapping to the same physical address) in the soft cache.
In particular, the Proxy Cache can invalidate the existing virtual address before responding to a load request for the same physical cacheline through a different virtual address.

\subsection{Control Hub} \label{sec:arch-control-tile}
The Control Hub consists of two submodules: the \textbf{FPGA Manager} and the \textbf{Soft Register Interface}.
The FPGA Manager provides necessary hardware support for programming and monitoring the eFPGA.
The programming engine loads the bitstream into the configuration memory, and performs integrity checks to detect data corruption.
The programmable clock generator either divides the system clock, or integrates a separate PLL for finer control over the generation of the FPGA clock.
The exception handler and the feature switches are similar to those in the Memory Hubs.
In particular, the feature switches can be used to set the timeout limit, reset the soft accelerator, or clear previously-logged error codes.

The Soft Register Interface enables the soft accelerator to implement a soft "device controller" similar to those commonly seen on off-chip peripheral devices.
The soft registers emulated by the eFPGA are accessible via on-chip MMIOs and often have additional effects rather than simply holding a value.
For example,
reading a soft register may dequeue from a FIFO inside the accelerator so that repetitive reads return different values.
When the Control Hub is deactivated, the Soft Register Interface returns bogus data to all processor accesses so that the system is not halted.

\subsection{Shadow Registers} \label{sec:arch-control-tile-shadow}

To prevent unwanted side effects, MMIOs typically adhere to a strict memory ordering model, e.g., I/O ordering.
Unfortunately, as stressed in Sec.~\ref{sec:arch-overview}, these strictly ordered accesses may stall the processors' pipelines because the eFPGA runs at a slower clock (Fig.~\ref{fig:sri-timing-normal}).
Addressing this inefficiency, the Soft Register Interface is augmented with several types of \textbf{Shadow Registers} residing in the fast clock domain (Fig.~\ref{fig:sri-timing-shadow}).
When a processor writes to a shadowed soft register, the Shadow Register acknowledges the request before forwarding the write into the eFPGA.
Conversely, the soft accelerator actively synchronizes shadowed soft registers over the asynchronous FIFO, so that the Shadow Registers can respond to processor reads immediately without notifying the eFPGA.

Duet supports four types of Shadow Registers: \textit{plain}, \textit{FPGA-bound FIFO}, \textit{CPU-bound FIFO}, and \textit{token FIFO}.
Plain Shadow Registers only keep the last value of multiple writes, which are ideal for passing constant parameters.
FPGA-bound and CPU-bound FIFOs, as their names suggest, record all writes from one side and allow the other side to read in order.
CPU-bound FIFO is \textit{blocking}, i.e., a processor read is stalled until the soft accelerator pushes into the FIFO or the request times out.
In contrast, CPU-bound \textit{token} FIFO is a dataless, \textit{non-blocking} FIFO that consumes a token or returns \textit{"empty"} in response to a processor read.
Token FIFO is designed particularly to emulate the non-blocking \texttt{try\_join} semantic in parallel programming models.


Note that normal soft registers are still available in case the software requires \textit{non-bufferable} accesses that must be sent all the way to the endpoint.
For example, a soft register can be dedicated as a barrier for synchronizing the processor and the eFPGA.
The processor signals its arrival at the barrier by reading the soft register, while the eFPGA signals its arrival at the barrier by acknowledging the read. 
To maintain I/O ordering, Shadow Register accesses are processed and responded to in order with respect to other shadowed or normal register accesses, as shown in Fig.~\ref{fig:sri-ordering}.

\input{figures/sri_timing}

%% file: figures/cache_options.tex
\begin{figure}[t!]
    \centering
    \hspace*{1.5cm}\includegraphics[scale=0.4, trim=0pt 3pt 0pt 3pt]{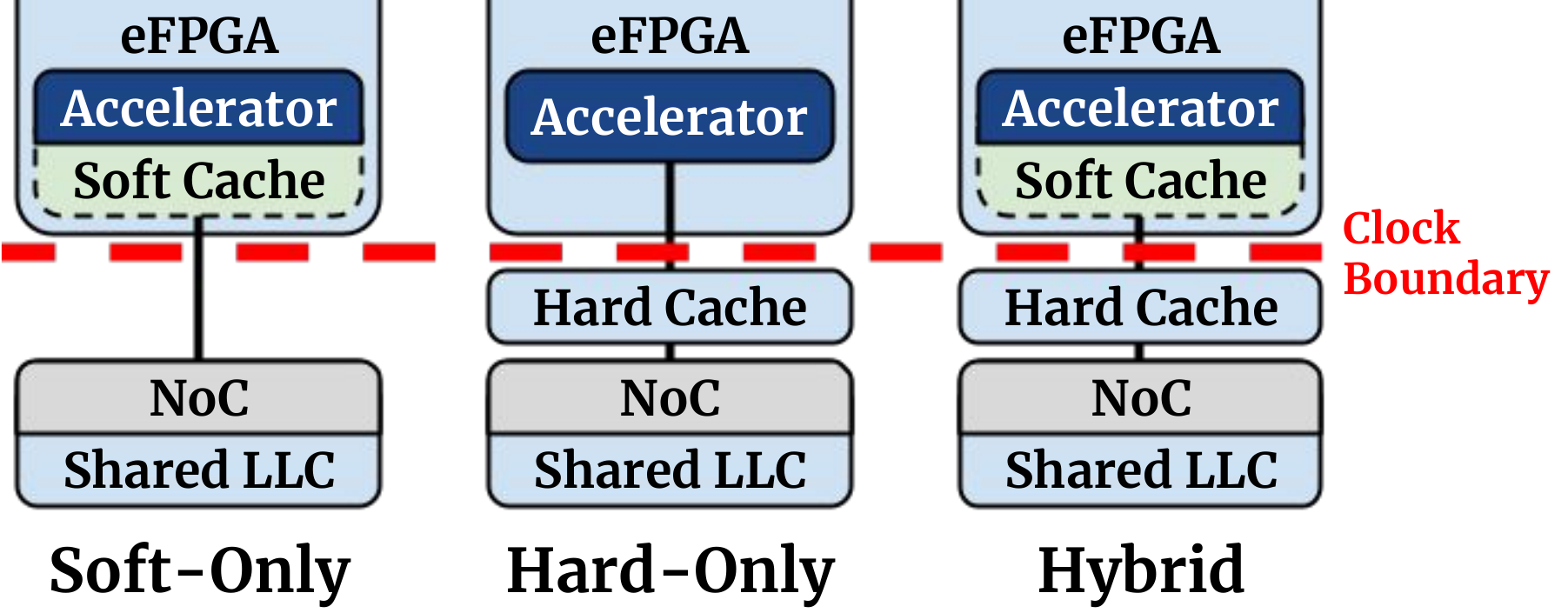}
    \setlength{\belowcaptionskip}{-4pt}
    \setlength{\abovecaptionskip}{-4pt}
    \caption{FPGA-Side Cache Organization Options}
    \label{fig:cache-options}
\end{figure}

%% file: figures/l2_timing.tex
\begin{figure}[!t]\centering

    
    \begin{subfigure}[t]{\linewidth}
        \centering
        \includegraphics[scale=0.4, trim=0pt 5pt 0pt 0pt, clip]{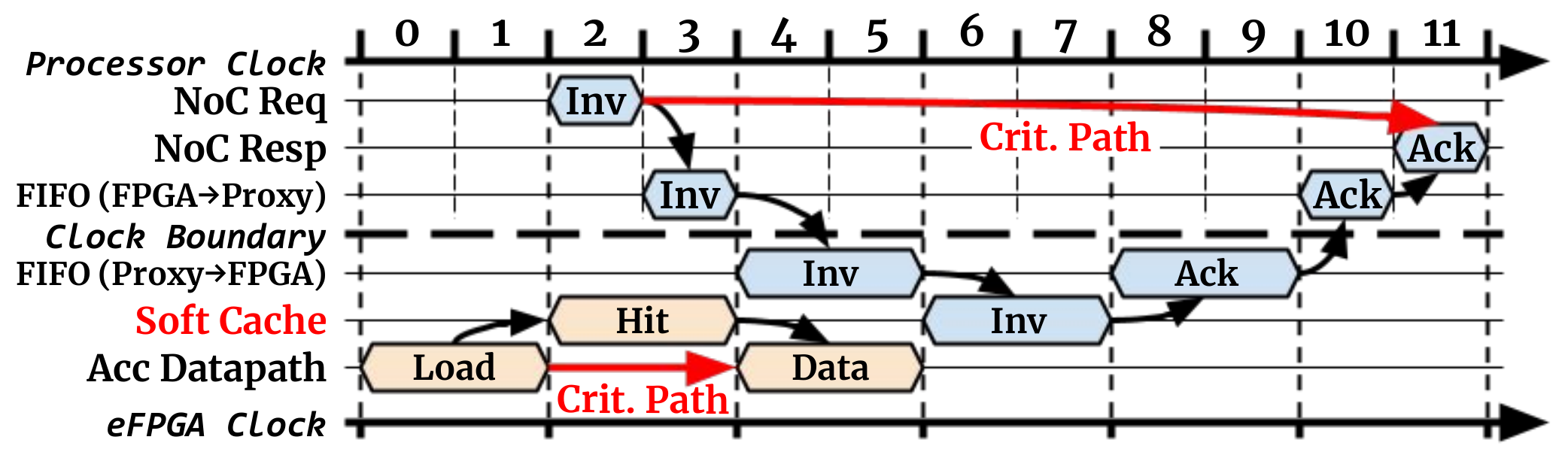}
        \caption{
        Cache Operations with Only \textbf{Soft} Caches
        }
        
        \begin{Center}
        \fontsize{9}{10}\selectfont
        \vspace{-0.6\baselineskip}
        \textbf{Inv}=Invalidation from a remote cache; \textbf{Ack}=Acknowledgement
        \end{Center}
        
        \label{fig:cache-soft-only}
    \end{subfigure}
    
    \begin{subfigure}[t]{\linewidth}
        \centering
        \includegraphics[scale=0.4, trim=0pt 5pt 0pt 0pt, clip]{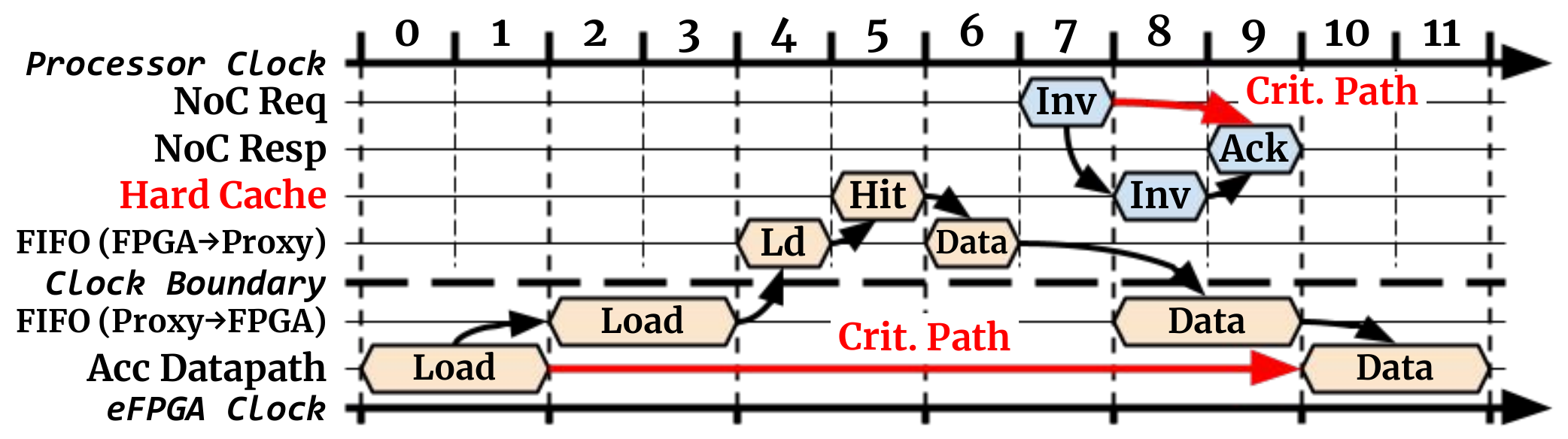}
        \caption{
        Cache Operations with Only \textbf{Hard} Caches}
        \label{fig:cache-hard-only}
    \end{subfigure}

    \begin{subfigure}[t]{\linewidth}
        \centering
        \includegraphics[scale=0.4, trim=0pt 5pt 0pt 0pt, clip]{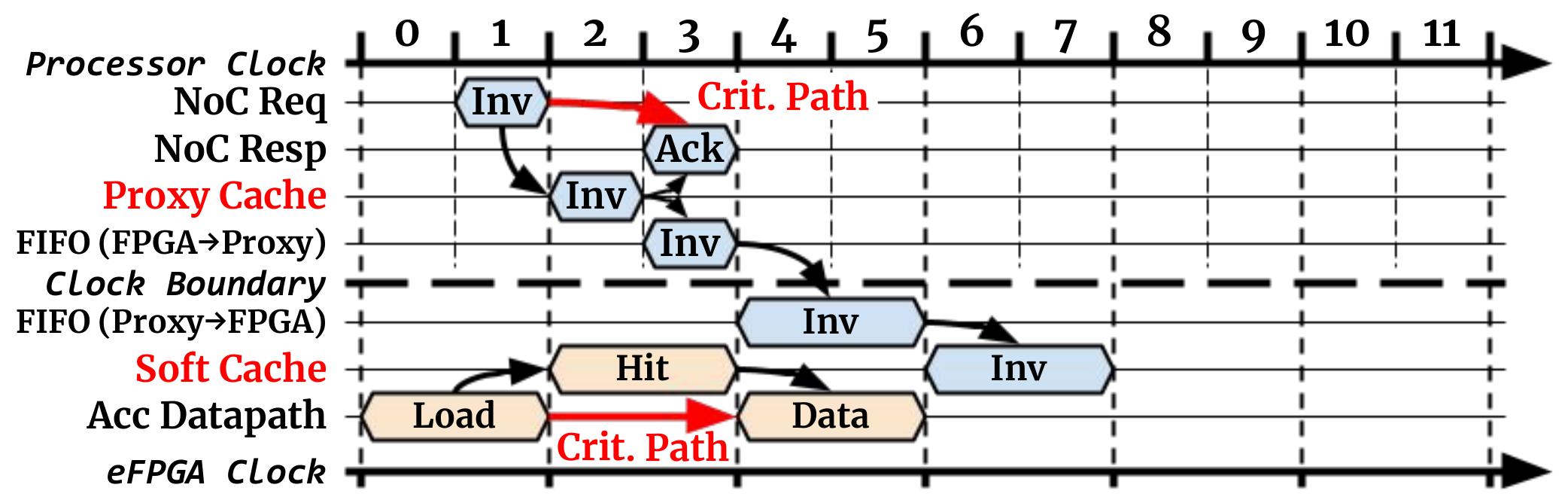}
        \caption{
        Cache Operations with Duet (This Work)
        }
        \label{fig:cache-duet}
    \end{subfigure}

    \caption{Cache Operations with Different Cache Organizations
    }
    
    \begin{FlushLeft}
    \fontsize{9}{9}\selectfont
    \vspace{-0.5\baselineskip}
    For simplicity, this figure shows single-stage async FIFOs, and the FPGA clock runs at half of the system clock frequency.
    On real hardware, async FIFOs typically take two to four stages, and the FPGA clock could be as slow as 1/10 of the system clock.
    \vspace{-1.5\baselineskip}
    \end{FlushLeft}
    \label{fig:cache-timing}
\end{figure}

%% file: figures/sri_timing.tex
\begin{figure}[!t]\centering
    
    \begin{subfigure}[t]{\linewidth}
        \centering
        \includegraphics[scale=0.40, trim={3pt 0 3pt 0}, clip]{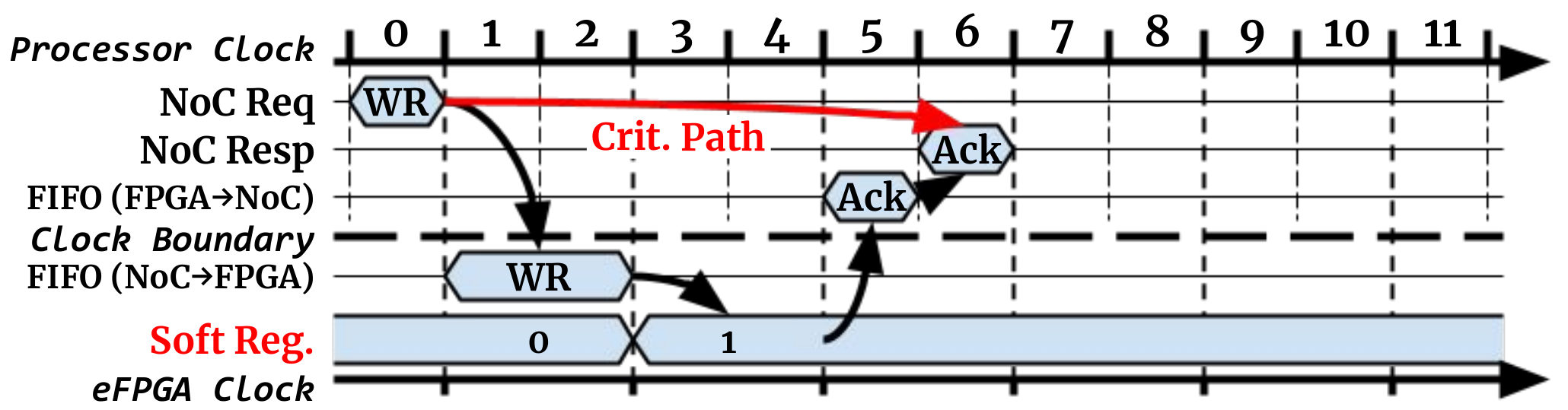}
        \setlength{\abovecaptionskip}{4pt}
        \caption{
        \fontsize{9}{10}\selectfont
        Accesses to a Normal Soft Register
        }
        \label{fig:sri-timing-normal}
    \end{subfigure}

    \begin{subfigure}[t]{\linewidth}
        \centering
        \includegraphics[scale=0.40, trim={3pt 0 3pt 0}, clip]{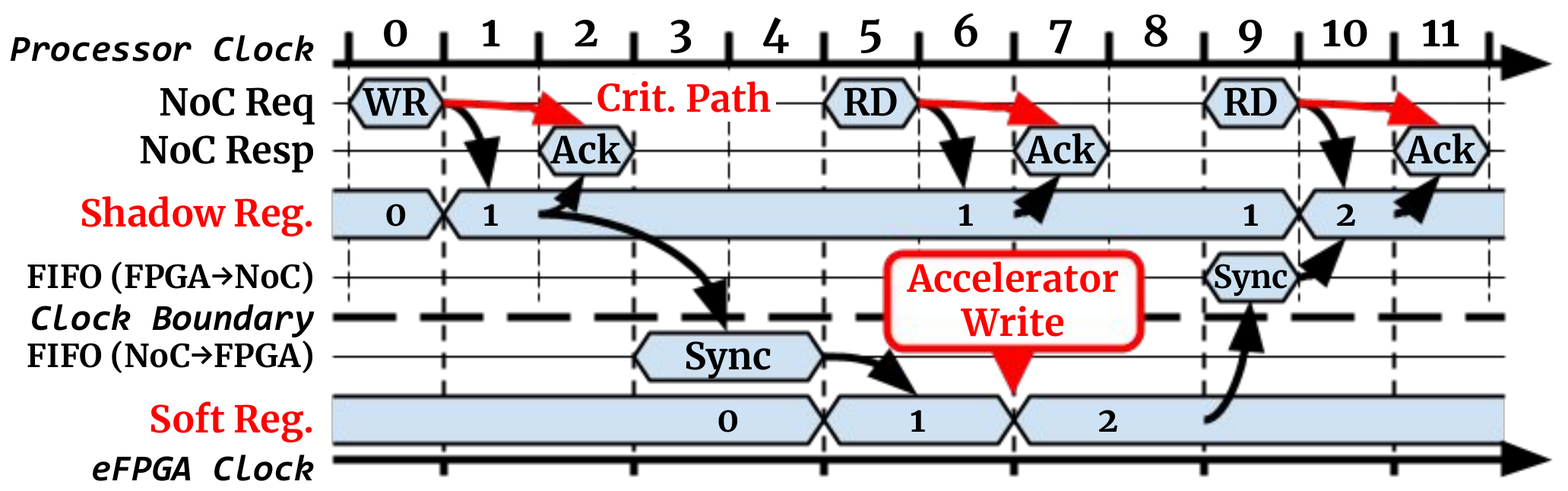}
        \setlength{\abovecaptionskip}{4pt}
        \caption{
        \fontsize{9}{10}\selectfont
        Accesses to a Shadowed Soft Register (This Work)
        }
        \label{fig:sri-timing-shadow}
    \end{subfigure}

    \begin{subfigure}[t]{\linewidth}
        \centering
        \includegraphics[scale=0.40, trim={3pt 0 4pt 0}, clip]{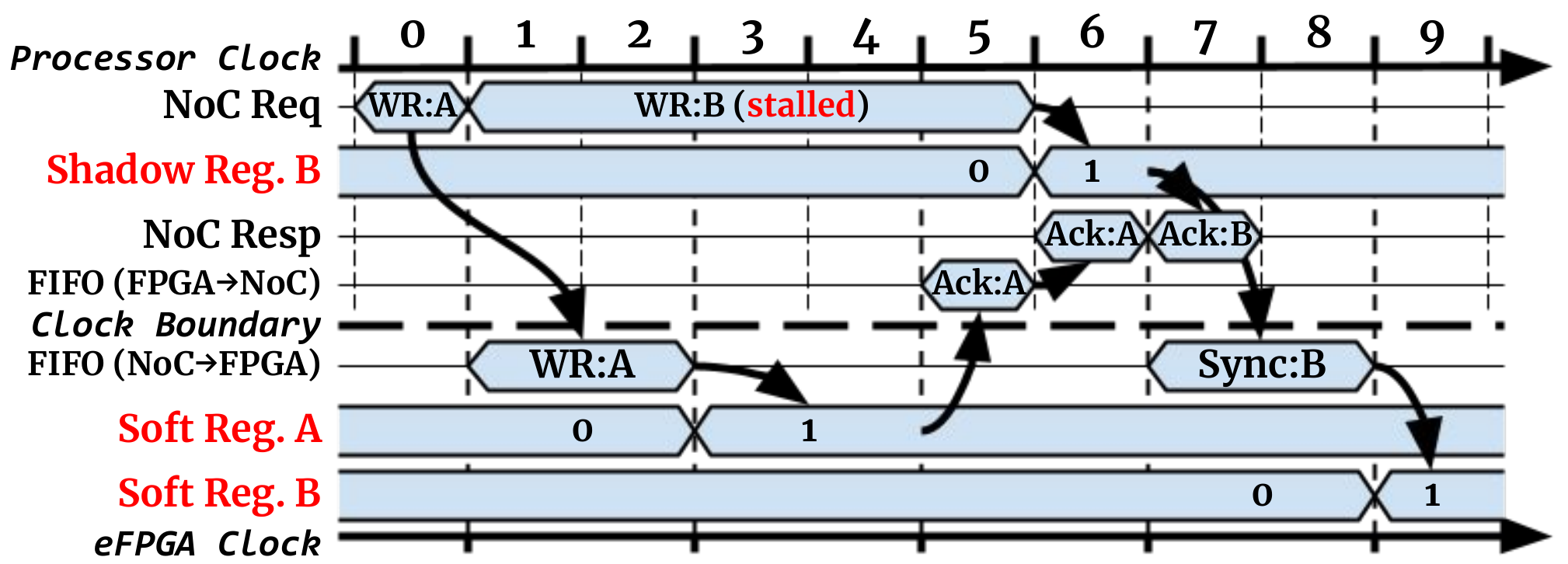}
        \setlength{\abovecaptionskip}{4pt}
        \caption{
        \fontsize{9}{10}\selectfont
        Strict Ordering of Two Soft Register Writes (This Work)
        }
        
        \label{fig:sri-ordering}
    \end{subfigure}

    
    \setlength{\belowcaptionskip}{-8pt}
    \setlength{\abovecaptionskip}{2pt}
    \caption{Accessing Soft Registers and Shadow Registers
    }
    
    \vspace{-0.5\baselineskip}
    \label{fig:sri-timing}
\end{figure}

%% file: tex/application.tex
\section{Applications} \label{sec:applications}

Duet
enables two novel paradigms of hardware acceleration, namely fine-grained acceleration and hardware augmentation.
In this section, we illustrate both paradigms with examples and discuss why Duet is the ideal architecture for them.

\subsection{Fine-Grained Acceleration} \label{sec:fine-grained-acc}

\subsubsection{Overview}
At its core, fine-grained acceleration and coarse-grained acceleration are similar because they both offload computation onto application-specific accelerators.
In fact, there is a spectrum rather than a clear boundary between the two paradigms \textemdash~the shorter the accelerated function is, or the more frequent the accelerator interacts with the processors, the closer it is to the fine-grained end of the spectrum.

Fine-grained acceleration has the following advantages:
first, it fully utilizes the compute power of the processors by running the less accelerable fractions of an algorithm on them, for example dynamic control flow, memory/IO-bound tasks, managing complex data structures, etc.;
second, it reduces software and accelerator design costs by moving less logic onto the FPGA;
third, in comparison to a monolithic accelerator, a collection of small accelerators are more composable, reusable, and more resilient to software updates.

\subsubsection{Example} \label{sec:barnes-hut-example}

Barnes-Hut (BH) algorithm~\cite{BH} is an approximation algorithm for 
simulating a dynamic system of particles
interacting via certain physical forces such as gravity.
In each time step, the simulation volume is divided into hierarchical, cubic cells stored in an octree (for three-dimensional space), so that the total force exerted by particles in distant cells can be approximated through a center-of-mass abstraction or a low-order multipole expansion.
Listing~\ref{lst:barnes-hut} shows the key functions that calculate the net force on all particles.

\input{listings/barnes-hut}

Prior work~\cite{BH-fpga} has proposed an FPGA-based, coarse-grained BH accelerator which contains replicated pipelines to parallelize force calculation on multiple particles.
However, despite the use of a special \textit{traversal cache} that facilitates data reuse between the pipelines, control flow divergence often leads to low pipeline utilization.
Furthermore, the accelerator relies on the host processor to serialize the BH-tree and preload the node pointers into the \textit{traversal cache}, making it inapplicable if the dataset exceeds the FPGA's BRAM capacity.
In summary, building a coarse-grained BH accelerator can be an intimidating task without a satisfactory result.

Fig.~\ref{fig:bh-fine-grained} shows the timeline of BH running on a dual-core system with two fine-grained accelerators.
Besides being easy to program, this software-hardware co-design is scalable and flexible:
multiple accelerators of each type can be instantiated to increase parallelism;
if \texttt{THRESHOLD} is high, i.e., \texttt{ApproxForce} is called for less times, the system can allocate more eFPGA resources to implement more and/or faster \texttt{CalcForce} accelerators, and vice versa if \texttt{THRESHOLD} is low.


The fine-grained BH accelerators would be inefficient without Duet's architectural support.
First, the accelerators access just a few cachelines at \textbf{random} memory locations per execution, resulting in poor utilization of the block-oriented, bandwidth-optimized memory system in most existing CPU-FPGA architectures.
In contrast, Duet's hybrid cache organization minimizes accelerators' memory access latency, especially for random accesses at cacheline granularity.
Second, the accelerators are invoked frequently and time-multiplexed by multiple CPU threads.
A centralized CPU-FPGA interconnect may be congested by the MMIOs issued by the processors to invoke the accelerators.
Duet's scalable integration makes it possible to split the workload across multiple eFPGAs,
and the shadow registers can further minimize MMIO latency.

\input{figures/bh_fine_grained}

\subsection{Hardware Augmentation}
\label{sec:hardware-aug}

\subsubsection{Overview}
Hardware augmentation allows application developers to add various hardware widgets to the system after chip fabrication.
These widgets help improve processor efficiency and/or mitigate software overheads in certain execution models, providing application-agnostic acceleration.
Different hardware augmentation widgets often exhibit unique behaviors and provide distinct software APIs.
For example, decoupled access-execute engines~\cite{dae} hide the latency of memory redirects by dereferencing software-specified pointers and collect the data into a hardware queue;
hardware garbage collectors~\cite{hw-gc} run autonomously in the background and alleviate software overhead of the runtime environment.
Since hardware augmentation widgets collaborate closely with the processors, Duet's scalable, tight, cache-coherent integration is critical to maximize performance.

\subsubsection{Example}
\label{sec:pdes}

Discrete event simulation (DES) is a well-known challenge for parallel execution~\cite{pdes}.
All the simulating threads must either progress in lock step (conservative method), which suffers from high synchronization overheads,
or exploit speculative execution (optimistic method), which requires a carefully designed software runtime or an architecture that supports thread-level speculation, e.g., SWARM~\cite{swarm}.

Duet opens up another possibility through hardware augmentation:
an eFPGA-emulated task scheduler can be loaded as part of the simulator software while offering hardware-level performance.
Consider a possible implementation of such a task scheduler.
Processors schedule new events by pushing memory pointers to the events into a \textit{FPGA-bound FIFO}, after which the task scheduler fetches the event data from shared memory and adds the pointer into the proper event queue.
Once certain events are ready to be processed, the task scheduler pushes the pointers into an \textit{CPU-bound FIFO} so that the processors can streamline event processing with minimal communication latency with the scheduler.
The task scheduler can support task speculation by fetching the cachelines that may be modified by a speculative event and storing versioned copies of them in its non-coherent memory.
On a mis-speculation, the task scheduler rolls back the cachelines to the most up-to-date, non-speculative versions, then reschedules the mis-speculated events.

%% file: listings/barnes-hut.tex
\begin{lstlisting}[
    language=c++,
    caption={The key compute loop of the Barnes-Hut simulation algorithm~\cite{BH}.
    The highlighted functions are static, compute-intensive, and ideal for fine-grained acceleration.
    },
    basicstyle=\fontsize{8}{8}\selectfont\ttfamily,
    commentstyle=\textcolor{gray},
    label={lst:barnes-hut},
    % numbers=left,
    % numberstyle=\tiny,
    % stepnumber=5,
    frame=trBL,
    % frame=tb,
    lineskip=-2pt,
    float={!tp},                    
    columns=fullflexible,
    keepspaces=true,
    escapechar=`,
    morekeywords={parallel_for},
    belowskip=-10pt
]
// Calculate the net force on particle "p"
void CalculateNetForce (BHTreeNode node, Particle p):
  if (Distance (node, p) > node.radius * THRESHOLD)
    // "p" is far enough from the particles in the subtree of
    // "node", so we approximate the net force with a low-order
    // multipole expansion and stop traversing the subtree
    p.force += `\hl{ApproxForce}` (node, p);
  else if (IsLeafNode (node))
    for (Particle q : node.children)
      p.force += `\hl{CalcForce}` (q, p);
  else
    // traverse the subtree via recursive calls
    for (BHTreeNode child : node.children)
      CalculateNetForce (child, p);

// Calculate the net forces on all particles in parallel
void CalculateNetForceOnAllParticles (
    BHTreeNode root, vector<Particle> allParticles):
  parallel_for (Particle p : allParticles)
    CalculateNetForce (root, p);
\end{lstlisting}


%% file: figures/bh_fine_grained.tex
\begin{figure}[t!]
    \centering
    \includegraphics[scale=0.54]{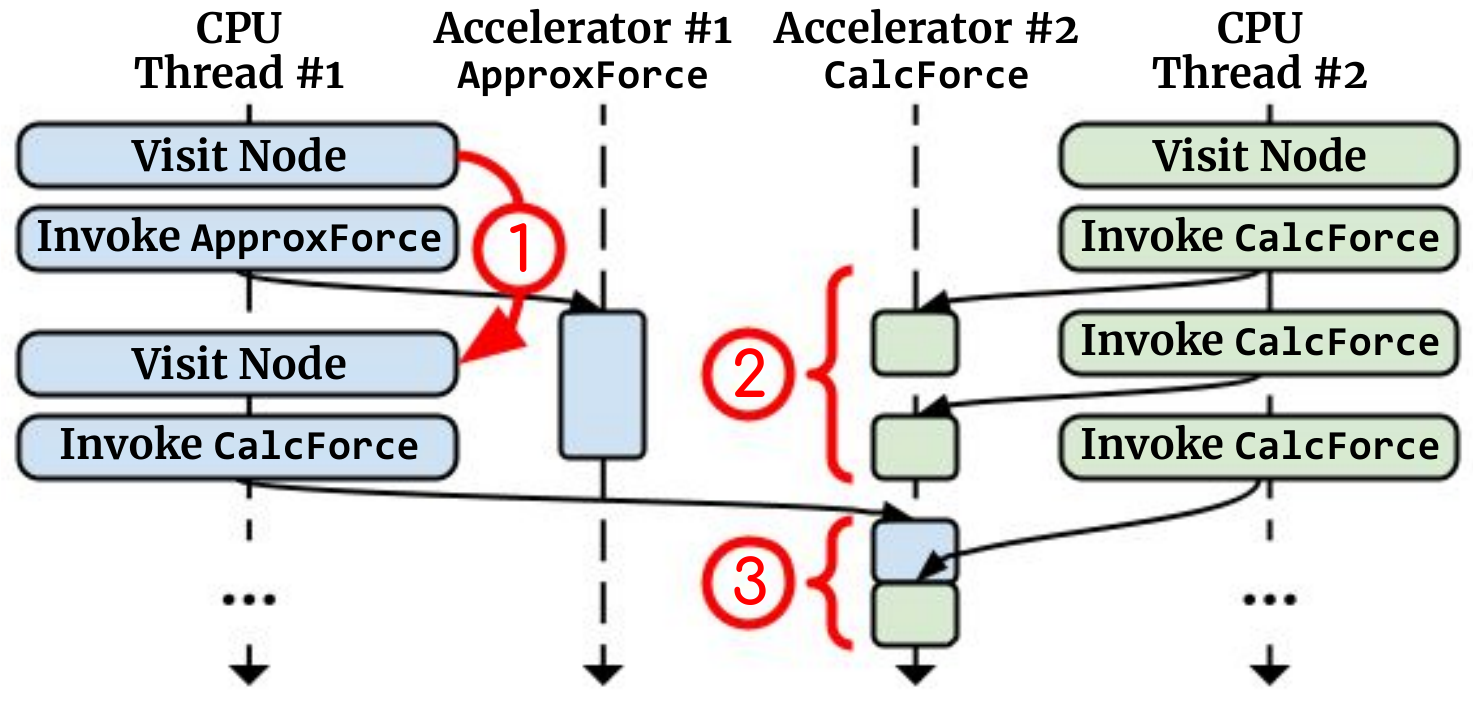}
    \setlength{\abovecaptionskip}{4pt}
    \caption{Multi-Threaded BH with Fine-grained Acceleration}
    
    \begin{FlushLeft}
    \fontsize{9}{9}\selectfont
    \vspace{-0.5\baselineskip}
    \circlenum{1} Loop-carry dependencies and dynamic control flow are handled by the processors;
    \circlenum{2} Processors and accelerators run in parallel through software pipelining;
    \circlenum{3} Accelerators are time-multiplexed by multiple CPU threads to increase utilization.
    \vspace{-1.5\baselineskip}
    \end{FlushLeft}
    
    \label{fig:bh-fine-grained}
\end{figure}

%% file: tex/dolly.tex
\section{Dolly}
\label{sec:instance}

\input{figures/instance_arch}

To evaluate Duet with high fidelity, we built Dolly\footnote{Named after \textit{Dolly Suite, Op. 56}, a collection of pieces for piano duet composed by Gabriel Fauré.}, a prototype system at the RTL level (Verilog/SystemVerilog) with a full toolchain for software development and accelerator design.
Dolly is configurable in the number of processors, the logic capacity of the eFPGA, the number of memory hubs, etc.
For simplicity, we name each Dolly instance P$p$M$m$, where $p$ specifies the number of processors, and $m$ specifies the number of memory hubs available to the eFPGA.
For example, Fig.~\ref{fig:instance-arch} shows the architecture of Dolly-P2M2.
\textbf{To encourage further research in tightly-integrated, hardware cache-coherent CPU-eFPGA systems, we have open-sourced Dolly and its toolchain, available at \url{https://github.com/PrincetonUniversity/Duet}}.

Dolly is based on the OpenPiton P-Mesh cache coherence system~\cite{openpiton} in a 2D mesh configuration.
The cache system is organized in three levels:
the L1 caches are tightly interwoven into the processors;
one private, write-back, 8KB L2 cache per physical tile interfaces with the corresponding L1 cache through the transaction-response interface (TRI)~\cite{byoc};
the shared L3 cache is distributed among all physical tiles, 64KB per shard, and runs a directory-based MESI protocol together with the private L2 caches.
The NoC offers point-to-point ordering of message delivery and supports additional message types besides the coherence messages, enabling on-chip MMIOs required by Dolly.

Dolly contains three types of physical tiles and an eFPGA.
Each \textit{P-Tile} hosts an Ariane~\cite{ariane} processor, a 6-stage, single-issue, in-order CPU which implements the 64-bit RISC-V instruction set.
Each Ariane processor has a private hardware FPU, an 8KB L1 instruction (L1I) cache, and an 8KB L1 data (L1D) cache.
\textit{C-Tiles} and \textit{M-Tiles} compose the Duet Adapter:
each \textit{C-Tile} contains one Control Hub (Sec.~\ref{sec:arch-control-tile}) and one Memory Hub (Sec.~\ref{sec:arch-memory-tile});
each \textit{M-Tile} includes one Memory Hub.
To demonstrate Duet's adaptability, Dolly implements the Proxy Cache by adding a \textit{coherent memory interface} to the \textit{unmodified} P-Mesh L2 cache.
The L2 cache, L3 cache shard and NoC router are the same across all physical tile types, so we wrap them into one module, the \textit{P-Mesh socket}.

The eFPGA is built with PRGA~\cite{prga} and employs a standard island-style architecture composed of logic blocks and routing blocks.
The eFPGA includes all of the common logic resources available on modern FPGAs, including look-up tables (LUT), bypassable flip-flops, hard adder chains, Block RAMs (BRAM), and hard multipliers.
Dolly employs a non-synthesizable clock generator to generate the eFPGA clock whose frequency can be specified in software.
All the asynchronous FIFOs are implemented with dual-clock RAMs and Gray-coded, 2-stage synchronizers. 



%% file: figures/instance_arch.tex
\begin{figure}[t!]
    \centering
    \includegraphics[width=\linewidth, trim=0pt 8pt 0pt 4pt]{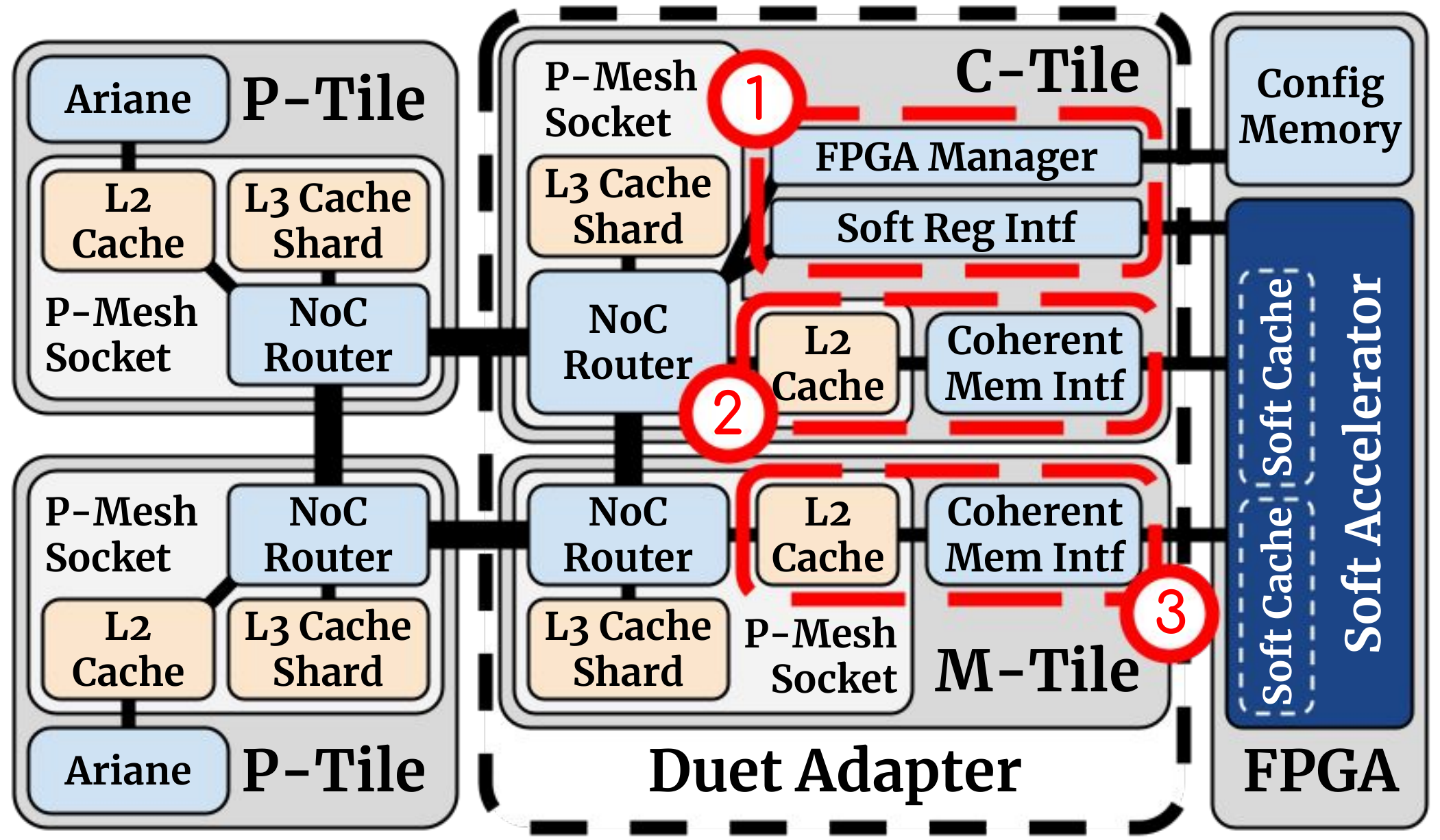}
    \setlength{\abovecaptionskip}{4pt}
    \caption{Architecture of Dolly-P2M2}
    
    \begin{FlushLeft}
    \fontsize{9}{9}\selectfont
    \vspace{-0.5\baselineskip}
    Dolly-P2M2 has 2 processors, 1 eFPGA, and 2 Memory Hubs.
    P-tile, C-tile and M-tile are physical tiles in a 2D mesh network.
    P-Mesh Socket is a physical wrapper for common components in all physical tiles, including an L2 cache, a NoC router, and a shard of the shared L3 cache.
    \circlenum{1} is the Control hub (Sec.~\ref{sec:arch-control-tile}).
    \circlenum{2} and \circlenum{3} are two Memory Hubs (Sec.~\ref{sec:arch-memory-tile}).
    Note that \circlenum{1} and \circlenum{2} reside in the same physical C-tile.
    \vspace{-1\baselineskip}
    \end{FlushLeft}
    \label{fig:instance-arch}
\end{figure}

%% file: tex/evaluation.tex
\section{Evaluation} \label{sec:eval}

%
%

\subsection{Overview} \label{sec:eval-overview}

We conduct our evaluation of Duet through RTL simulations of Dolly instances and focus on three perspectives:
first of all, we report the area consumption and the typical frequency of the hard components of Dolly (Sec.~\ref{sec:eval-arch});
second, we run synthetic benchmarks to measure the CPU-FPGA communication latency and bandwidth achieved by the Duet Adapter under different conditions (Sec.~\ref{sec:eval-itx});
third, we run seven application benchmarks and compare their performance
against processor-only and FPSoC baselines (Sec.~\ref{sec:eval-app}).

\textbf{To show the lower bound of Duet-enabled improvements, we give
the processors various advantages throughout our evaluation.}
Specifically, we boost the clock frequency of the Ariane processors and the P-Mesh cache system to 1GHz, despite that previous works reported their maximum frequencies (after scaling to 45$nm$) to be 455MHz and 711MHz, respectively.
For the application benchmarks, processor-only baselines always start with a warm cache, while the soft accelerators always start with a cold cache.

\subsection{Area and Typical Frequency of the Hard Components} \label{sec:eval-arch}

\input{tables/area}

Table~\ref{tab:area} summarizes the area and typical frequency of each hardware component of Dolly.
The numbers for Ariane and P-Mesh components are retrieved from previous works.
The submodules of the Control Hub and the Memory Hub are synthesized using Synopsys Design Compiler and Silvaco's Open-Cell Library~\cite{nangate45} based on the NCSU FreePDK45 process design kit~\cite{freepdk45} using an area utilization rate of 70\%.
In summary, the Control Hub and the Memory Hub require minimal hardware resources.
The area consumption of the eFPGA and the clock frequency of the emulated accelerator are reported on a per-benchmark basis in Sec.~\ref{sec:eval-app}.

\subsection{CPU-eFPGA Communication Latency and Bandwidth} \label{sec:eval-itx}

In this section, we evaluate the peak performance of Duet with a synthetic benchmark.
The eFPGA emulates a simple scratchpad memory and a processor uses different mechanisms to access it.
To show that the eFPGA's clock frequency has a significant impact on performance, we sweep the eFPGA clock from 20MHz to 500MHz while fixing the system clock at 1GHz.
In particular, we study the following CPU-eFPGA communication mechanisms and show how Duet improves them over existing CPU-FPGA systems:

Soft registers are memory-mapped, eFPGA-emulated registers which the processor can read and write via non-coherent MMIOs.
These soft registers can be implemented inside the eFPGA (\textbf{Normal Registers}), in which case processor accesses must pay the Clock-Domain-Crossing (CDC) overhead.
\textbf{Shadow Registers} address this inefficiency.
In this study, we use FPGA-bound FIFOs to handle processor writes and CPU-bound FIFOs for processor reads.

Alternatively, the processor can store the data in shared memory and pass the memory address to the eFPGA via a soft register write.
The soft register write not only signals the eFPGA to load the data (\textbf{eFPGA Pull}), but also works as a synchronization to ensure that all processor stores are committed into the cache system before the eFPGA starts loading.
The eFPGA can send data to the processor in a similar way (\textbf{CPU Pull}).
As described in Sec.~\ref{sec:arch-proxy-cache}, commodity FPSoCs typically emulate FPGA-side caches using eFPGA resources (\textbf{Slow Cache}), while Duet employs the novel \textbf{Proxy Cache} to improve cache performance.


\input{figures/synth_latency}

\textbf{Latency Study}

We first measure the minimum round-trip latency of the aforementioned communication mechanisms on Dolly-P1M1.
Fig.~\ref{fig:synth-latency} shows the breakdown of the CPU-eFPGA communication latency into four parts: NoC latency, cache processing time in the fast clock domain, cache processing time in the slow clock domain, and the CDC overhead.
All numbers are collected in an ideal scenario, that is, single processor (no contention or NoC congestion), single transaction (no buffer clogging).
Note that the eFPGA pulls and CPU pulls are guaranteed to miss in the requesting cache and to hit in the other party's private cache in a modified state.
Both the NoC latency and the cache processing time in the fast clock domain include the time for the distributed directory to send and process the secondary write-back requests. 

As we can see, the CDC overhead and the slow clock penalty on cache processing constitute over half of the round-trip latency for a slow cache, even when the eFPGA runs at 50\% of the CPU clock frequency.
For eFPGA pulls, the Proxy Cache can reduce the latency by 13\% to 43\%, and the reduction increases as the eFPGA runs slower.
For CPU pulls, the Proxy Cache achieves a constant latency regardless of the eFPGA clock frequency, reducing the latency by 42\% to 82\%.
The Shadow Registers also have a fixed latency, reducing CPU-eFPGA communication latency by 50\% to 80\%.
Note that the Proxy Cache and the Shadow Registers only move the eFPGA off of the critical path, and the eFPGA still needs time to issue memory accesses and to handle soft register accesses.

\textbf{Single-Processor Bandwidth Study}

\input{figures/synth_bw}

Next, we study the maximum bandwidth of single-processor communications on Dolly-P1M1.
In this study, we configure the synthetic benchmark to pass 512 quad-word (8 Bytes) integers from one processor to the eFPGA and then fetch them back.
With the soft registers, this is done by having the processor execute a loop that writes/reads one integer per iteration.
When using shared memory, the processor first allocates two 4KB buffers in the memory and passes the base addresses to the eFPGA using two plain shadow registers.
Then, the processor stores all the integers into one of the two buffers and sends a \textit{read} request to another normal soft register, awakening the eFPGA and blocking itself.
The eFPGA proactively loads the entire array into its scratchpad memory, stores it back to the other buffer, and then unblocks the processor by acknowledging the processor's read request.
Finally, the processor loads the array from shared memory.

Fig.~\ref{fig:synth-bw} shows the results of this study.
The Proxy Cache delivers the highest bandwidth across all eFPGA frequencies.
In fact, the Proxy Cache reaches the peak bandwidth for eFPGA pulls (558MB/s) when the eFPGA runs faster than 100MHz (10\% of the CPU clock frequency);
for CPU pulls, the peak bandwidth (201MB/s) is reached at 50MHz.
In comparison, the slow cache can only achieve 287MB/s for eFPGA pulls and 144MB/s for CPU pulls even when the eFPGA runs at 500MHz. 
The largest bandwidth gap is 9.5x when the eFPGA runs at 100MHz.
The upper bounds of cache-based communications are determined by the NoC bandwidth and the number of concurrent, in-flight, memory requests supported by the Proxy Cache.
Note that the upper bounds of CPU pulls are much lower than those of eFPGA pulls.
This is because the cache line size is 16 Bytes and the eFPGA can load up to one line per cycle,
but the L2 cache in Dolly only supports stores up to 8 Bytes, so the eFPGA must send two requests to store one cacheline, under-utilizing the asynchronous FIFOs.

The Shadow Registers also achieve a stable bandwidth (213MB/s) once the eFPGA clock frequency exceeds 10\% of the CPU clock frequency.
In comparison, normal registers can only reach 121MB/s at 500MHz.
The upper bound of soft register accesses is limited by the strict consistency model of MMIOs.
If the I/O consistency model is relaxed, the upper bound is then limited by the number of concurrent, in-flight MMIOs supported by the load-store queue of the processor.

\textbf{Multi-Processor Scalability Study}

One key difference between Duet and existing CPU-FPGA systems is the scalability.
At the architecture level, Duet has no constraint on the number of processors, eFPGA fabrics, or the number of Memory Hubs per eFPGA.
In this section, we study the effect of multi-processor contention on the soft registers.
In particular, we run the synthetic benchmark on multiple Dolly instances with different number of processors which constantly writes/reads the same soft register.
The eFPGA clock is fixed at 500MHz (50\% of the CPU clock frequency).

Fig.~\ref{fig:synth-nprocs} shows the results.
The Shadow Registers can stably support up to 8 processors before the per-processor bandwidth starts to drop, while the normal registers can only support up to 2 processors.
Considering that the processors need to perform other tasks between soft register accesses, the Shadow Registers can support even more processors in real applications.

\textbf{Summary}

In summary, the significant reduction in latency and the stable increase in bandwidth clearly justify the promotion of soft registers and private caches into the fast clock domain.

\input{figures/synth_nproc}
\subsection{Application Benchmarks} \label{sec:eval-app}
\input{figures/speedup}


In this section, we evaluate Duet with seven application benchmarks.
For each benchmark, we first run a C implementation as the \textbf{processor-only} baseline and record the total runtime.
Note that all benchmarks are run in bare metal due to limited simulation speed at the RTL level.
We then design the soft accelerators for each benchmark, either directly at the RTL level, or by synthesizing an alternative C implementation with Catapult HLS~\cite{mg-catapult}.
Each accelerator is synthesized, placed and routed with the PRGA~\cite{prga} workflow to get the accurate eFPGA clock frequency and an area estimation of the utilized FPGA resources.
In particular, we map each benchmark onto the flagship FPGA model provided by VTR~\cite{vtr} (\texttt{\footnotesize k6\_frac\_N10\_frac\_chain\_mem32K\_40nm}) which resembles an Altera Stratix IV FPGA.
With the accurate eFPGA clock frequency, we rerun the benchmarks on various \textbf{Dolly} instances and an \textbf{FPSoC}-like architecture, recording the total runtime which includes all the overhead in CPU-FPGA communication and synchronization.
In particular, the FPSoC model moves the P-Mesh L2 cache into the eFPGA's (slow) clock domain and downgrades all shadowed soft registers to normal registers.
We then compute the speedup over processor-only baselines and compare the results achieved by Dolly and the FPSoC model.

We use Area-Delay-Product (ADP) to quantify and compare area efficiency.
When calculating area consumption, the processor-only baseline only counts the processors and the hardware cache system.
The FPSoC-like architecture adds the silicon area of the FPGA on top of the processor-only baseline, and Dolly further includes the Duet Adapters.

The seven benchmarks, their corresponding Dolly instance, and the acceleration paradigm (\textbf{FG} for fine-grained acceleration, \textbf{HA} for hardware augmentation) are the following:

\begin{itemize}[leftmargin=0em, itemsep=0pt, itemindent=1em]
    \item \textbf{\texttt{Tangent}} (P1M0, FG):
    A floating-point Tangent accelerator is implemented with Catapult HLS using a piece-wise linear approximation algorithm with a maximum error rate of 0.3\% compared to the C math library (\texttt{libm}).
    An FPGA-bound FIFO is used to pass the argument to the accelerator and invoke it.
    Results are returned through an CPU-bound FIFO.
    
    
    \item \textbf{\texttt{Popcount}} (P1M1, FG): 
    Popcount counts the number of ones in a long bit vector (512 bits).
    Since the Ariane processor does not support the RISC-V BitManip Extension, we use a byte look-up algorithm for the processor-only baseline.
    The accelerator is hand-written in Verilog and uses one Memory Hub to load the bit vector from coherent memory.
    
    \item \textbf{\texttt{Sort}} (P1M2, FG):
    We use the SPIRAL Project~\cite{spiralsort} to generate 3 sorting networks in Verilog for sorting 32, 64, 128 double-word (4-Byte) integers.
    The accelerator uses two memory hubs, one for reading the input array from coherent memory and one for writing the sorted array back, so that the accelerator can be pipelined to sort fixed-length slices of a larger array which can then be merge-sorted by the processor. 
    The processor-only baseline runs quicksort on the entire array.
    
    \item \textbf{\texttt{Dijkstra}} (P1M1, FG):
    We implement an accelerator for Dijkstra's Shortest Path algorithm with Catapult HLS and use a soft cache to exploit data locality between consecutive calls to the accelerator.
    
    \item \textbf{\texttt{Barnes-Hut}} (P4M1, FG):
    As explained in Sec.~\ref{sec:barnes-hut-example}, the two key kernels in the Barnes-Hut algorithm are implemented as soft accelerators using Catapult HLS.
    Both accelerators are pipelined and time-multiplexed by four processors.
    The processor-only baseline also parallelizes force calculation for different particles across four processors.
    
    
    \item \texttt{\textbf{PDES}} (P4M1/P8M1/P16M1, HA):
    As described in Sec.~\ref{sec:pdes}, a non-speculative, hardware task scheduler is designed in Verilog
    to accelerate Parallel Discrete Event Simulation (PDES) of a digital circuit.
    The processor-only baseline uses MCS locks~\cite{mcs-lock} to arbitrate accesses to the shared event queue, and the lock contention can be severe as the number of cores increases.
    
    \item \textbf{\texttt{BFS}} (P4M0/P8M0/P16M0, HA):
    We implement multiple hardware, lock-free queues in Verilog to alleviate the synchronization overhead in parallel Breadth-First Search (BFS).
    The processors traverse the graph in barrier-synchronized steps and use the queues to store the current and next search frontiers.
    Similar to the PDES benchmark, the processor-only baseline suffers from synchronization bottlenecks.
    
\end{itemize}

\input{tables/app_util}

Table.~\ref{tab:eval-app} lists the maximum clock frequency and eFPGA resource utilization of the soft accelerators.
Note that the soft accelerators can only run at 8\% - 28\% of the processors' clock frequency.
As studied in Sec.~\ref{sec:eval-itx}, Duet achieves peak bandwidth in this frequency range and outperforms the conventional FPSoC communication mechanisms by at least 2x.

Fig.~\ref{fig:speedup} shows the normalized speedup and ADP of the application benchmarks.
Duet outperforms FPSoC across all benchmarks and achieves a geometric mean of 4.53x speedup over processor-only baselines, in comparison to 2.14x achieved by FPSoC using the same accelerators.
The sorting accelerators provide up to 16.2x speedup on Duet but only 4.0x on FPSoC, because the FPGA-side caches in FPSoC are penalized by the slow clock cycles.
Some may argue that the FPGA-side caches are unnecessary because sorted arrays are consecutive blocks for which DMAs are efficient.
However, due to the small array sizes (128-512B), even LLC-coherent DMA incurs too much control overhead in cache flushing.
The lock-free queues used in BFS provide up to 24.9x speedup on Duet but only 7.8x on FPSoC, because the processor-accelerator communication latency on FPSoC is much higher without the shadow registers.
Note that the BFS benchmark shows a superlinear speedup when scaling from a 4-core system (Dolly-P4M0) to an 8-core system (Dolly-P8M0).
This is because the processor-only baseline sees a performance decrease when scaling from 4 cores to 8 cores due to intensifying lock contention.

In terms of area efficiency, Duet achieves a geometric mean of 0.39x lower ADP than the processor-only baseline (lower is better), while the FPSoC is 0.23x higher.
Duet outperforms FPSoC on most benchmarks except for \texttt{Dijkstra}.
This is because the FPSoC model hardens the FPGA-side cache in the slow clock domain, so that soft caches become unnecessary and can be removed to save eFPGA resources.

In summary, Dolly achieves superior speedup with the same eFPGA-emulated accelerators than FPSoCs.
Therefore, we can conclude that Duet provides better support for fine-grained acceleration and hardware augmentation.

%% file: tables/area.tex
\begin{table}[t]

    \fontsize{8}{8}\selectfont
    \setlength\tabcolsep{1.5pt}
    \centering
    \newcounter{notenumber}
    \newcommand{\mynote}[1]{\setcounter{notenumber}{#1}\textsuperscript{\fnsymbol{notenumber}}}

        \caption{\centering
        Area and Typical Frequency of Dolly Components}
        
    \begin{threeparttable}[t]
        \begin{tabular}{|c|c|c|c|c|c|}
        
        \hline
        
        \multirow{2}{*}{Component} &
        Device &
        Area &
        Freq. &
        \textbf{Scaled Area}\mynote{1} &
        \textbf{Scaled Freq.}\mynote{1} \\
        
        &
        Technology &
        (mm\textsuperscript{2}) &
        (MHz) &
        \textbf{(mm\textsuperscript{2})} &
        \textbf{(MHz)} \\
        
        \noalign{\hrule height 1.5pt}
        
        \multirow{2}{*}{Ariane~\cite{ariane-area}} &
        GlobalFoundries &
        \multirow{2}{*}{0.39} &
        \multirow{2}{*}{910} &
        \multirow{2}{*}{\textbf{1.56}} &
        \multirow{2}{*}{\textbf{455}\mynote{2}} \\
        
        &
        22$nm$ FDX &
        &
        &
        &
        \\
        
        \hline
        
        P-Mesh &
        \multirow{2}{*}{IBM~32$nm$~SOI} &
        \multirow{2}{*}{0.55} &
        \multirow{2}{*}{1000} &
        \multirow{2}{*}{\textbf{1.1}\mynote{1}} &
        \multirow{2}{*}{\textbf{711}\mynote{1}\mynote{2}} \\
        
        Socket~\cite{openpiton} &
        &
        &
        &
        &
        \\
        
        \hline

        FPGA~Mgr~+ &
        \multirow{4}{*}{FreePDK45} &
        \multirow{2}{*}{0.21} &
        \multirow{2}{*}{925} &
        \multirow{2}{*}{\textbf{0.21}\mynote{1}} &
        \multirow{2}{*}{\textbf{925}\mynote{1}} \\
        
        Soft~Reg~Intf &
        &
        &
        &
        &
        \\
        
        \cline{1-1} \cline{3-6}
        
        Coherent &
        &
        \multirow{2}{*}{0.04} &
        \multirow{2}{*}{1250} &
        \multirow{2}{*}{\textbf{0.04}\mynote{1}} &
        \multirow{2}{*}{\textbf{1250}\mynote{1}} \\
        
        Memory Intf &
        &
        &
        &
        &
        \\
        
        
        
        
        \hline
        \end{tabular}
        
        \setcounter{notenumber}{0}
        
        \begin{tablenotes}
            \item [\stepcounter{notenumber}\fnsymbol{notenumber}] \fontsize{9}{9}\selectfont Scaled to 45$nm$ with a linear MOSFET scaling model
            \item [\stepcounter{notenumber}\fnsymbol{notenumber}] \fontsize{9}{9}\selectfont \rebut{To emulate a higher-performance multi-core, our evaluation simulates the Ariane cores and the hardware cache system at 1GHz}
        \end{tablenotes}
        
        
    \end{threeparttable}
    \label{tab:area}

    \end{table}

%% file: figures/synth_latency.tex
\begin{figure}[t!]
    \centering
    \includegraphics[width=\linewidth, trim=0pt 10pt 0pt 8pt]{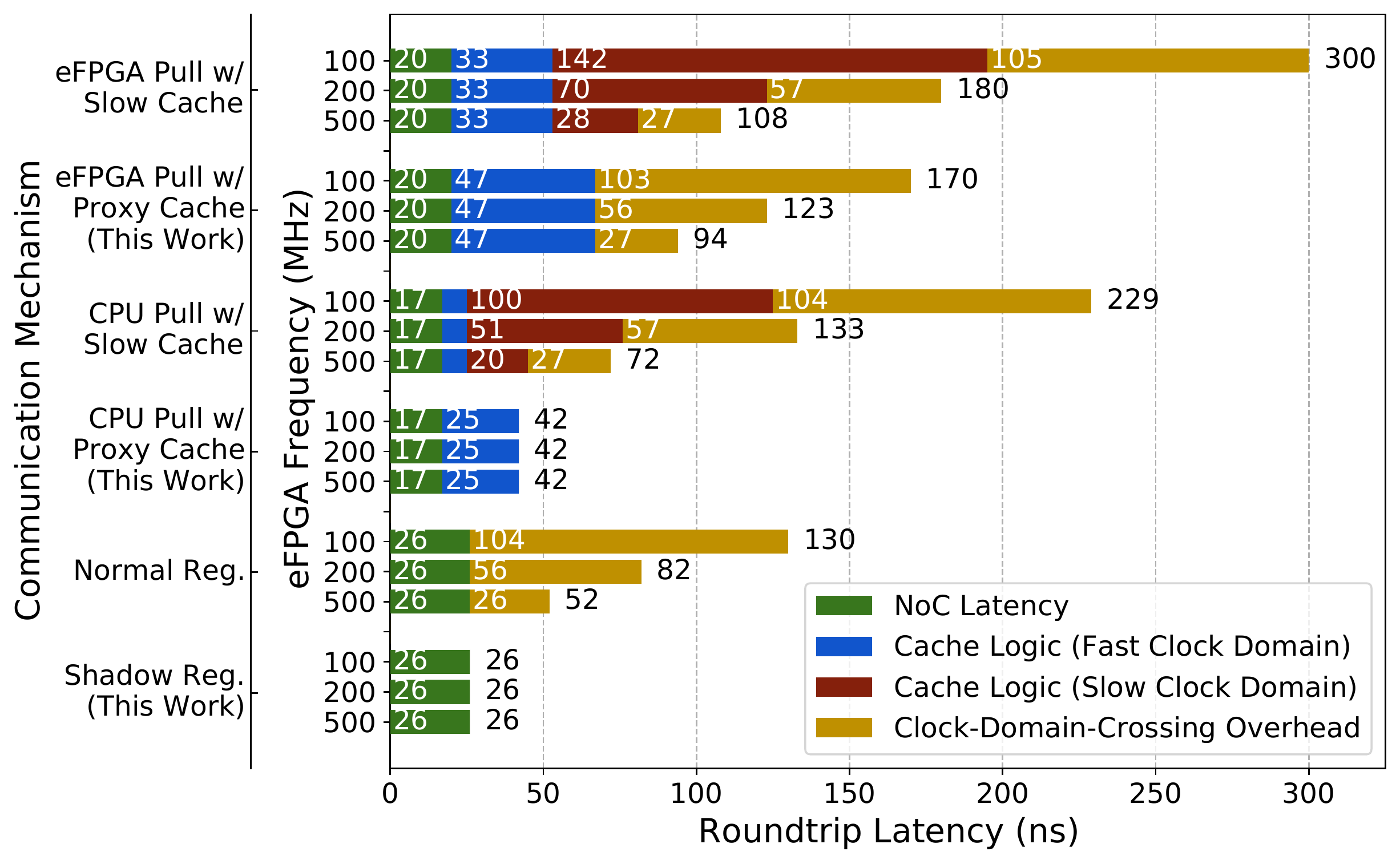}
    \setlength{\belowcaptionskip}{-8pt}
    \setlength{\abovecaptionskip}{-4pt}
    \caption{\centering 
    CPU-eFPGA Communication Latency \newline
    \fontsize{9}{10}\selectfont \normalfont (Single processor; Single transaction; Lower is better)\vspace{0\baselineskip}}
    
    
    \label{fig:synth-latency}
\end{figure}

%% file: figures/synth_bw.tex
\begin{figure}[t!]
    \centering
    \includegraphics[width=\linewidth, trim=0pt 10pt 0pt 8pt]{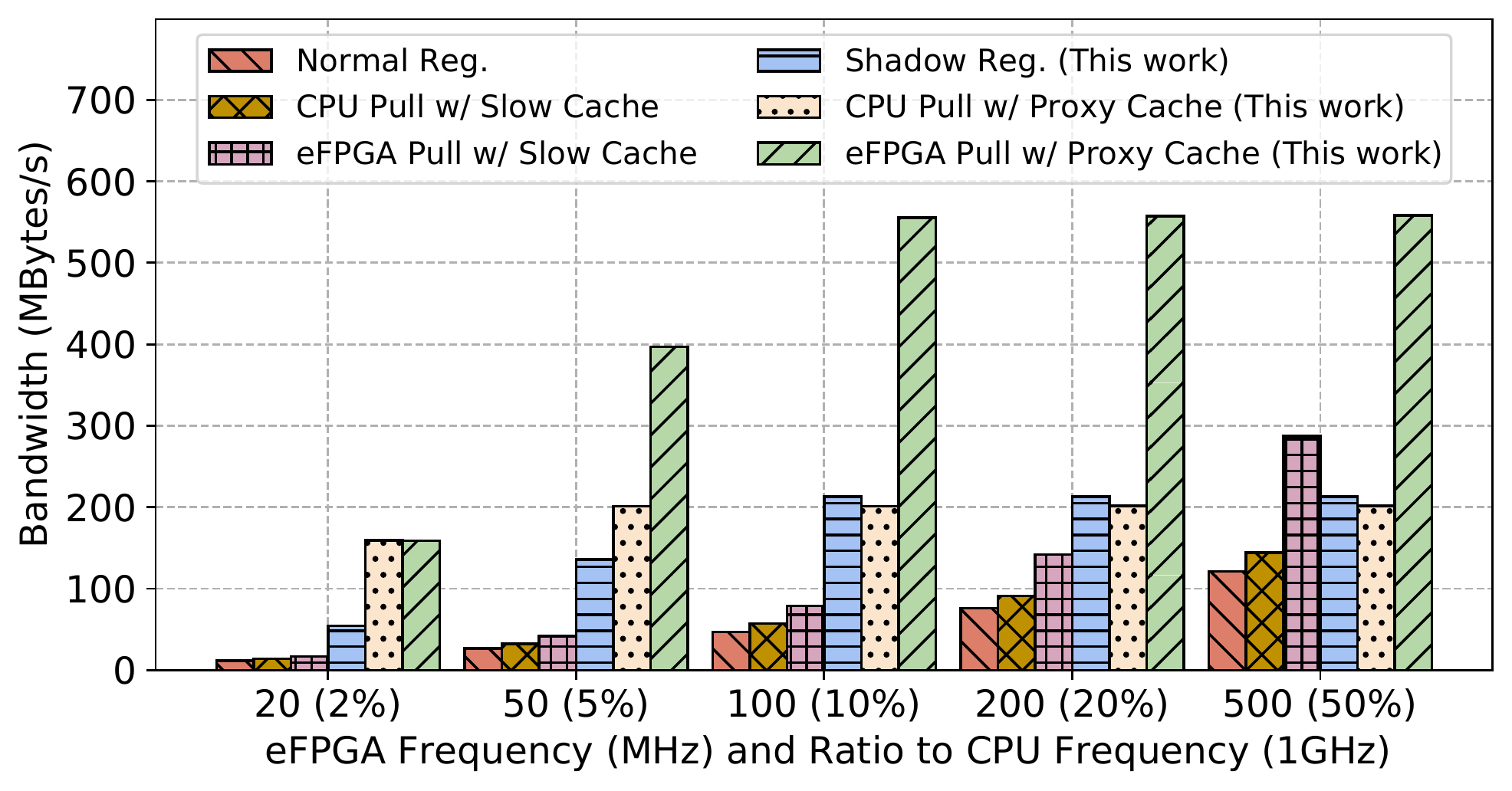}
    \setlength{\belowcaptionskip}{-4pt}
    \setlength{\abovecaptionskip}{4pt}
    \caption{
    Processor-eFPGA Communication Bandwidth vs. eFPGA Clock Frequency
    {\fontsize{9}{\baselineskip}\selectfont \normalfont (Single processor; Higher is better)}
    }
    
    \vspace{-0.5\baselineskip}
    
    \label{fig:synth-bw}
\end{figure}

%% file: figures/synth_nproc.tex
\begin{figure}[t!]
    \centering
    \includegraphics[width=\linewidth, trim=0pt 10pt 0pt 8pt]{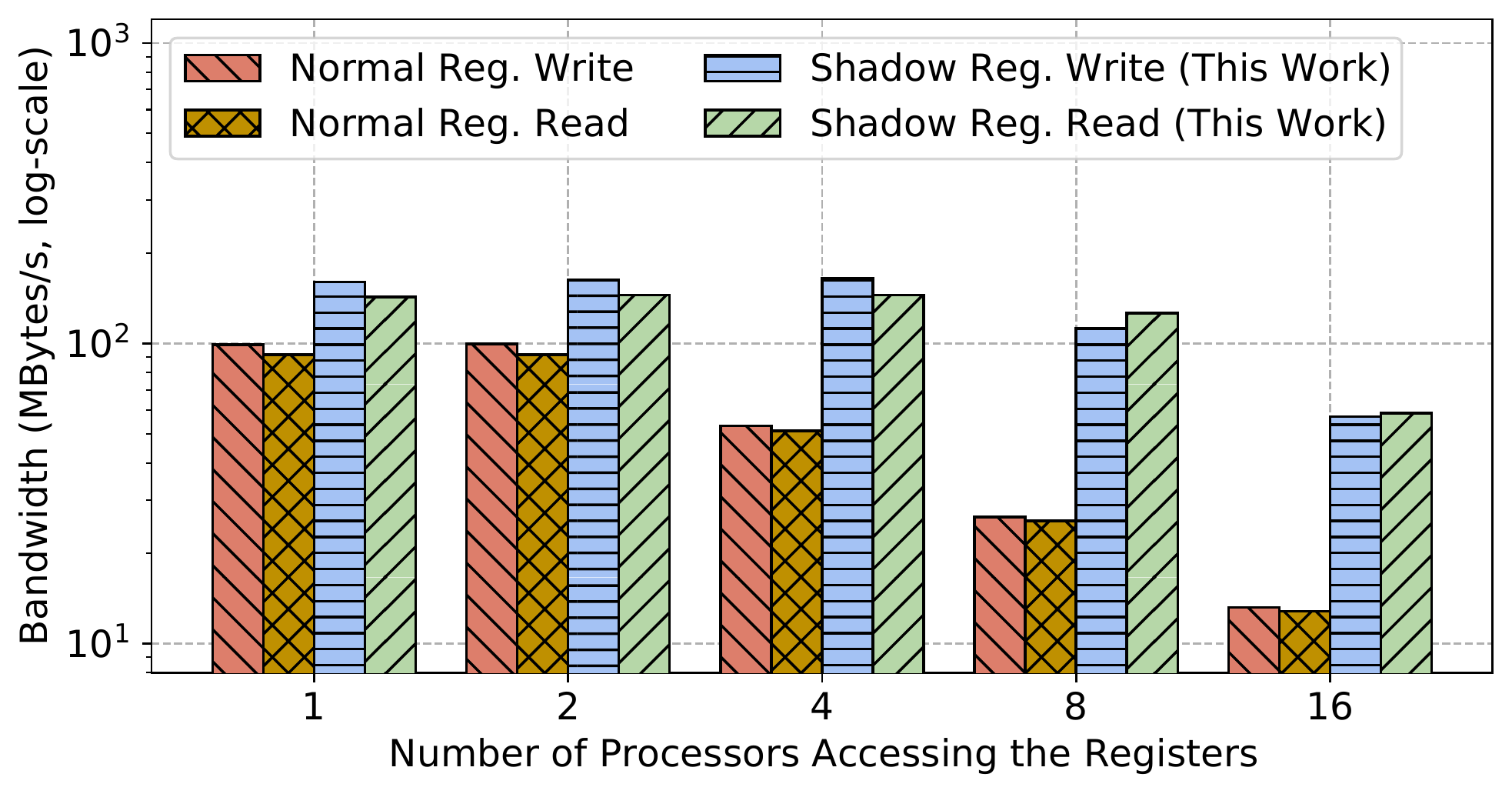}
    \setlength{\belowcaptionskip}{-8pt}
    \setlength{\abovecaptionskip}{4pt}
    \caption{\centering 
    Processor-eFPGA Communication Bandwidth (Per Processor) vs. Number of Contending Processors}
    
    \label{fig:synth-nprocs}
\end{figure}

%% file: figures/speedup.tex
\begin{figure*}[t!]
    \centering
    \includegraphics[scale=0.40, trim=0pt 2pt 0pt 6pt]{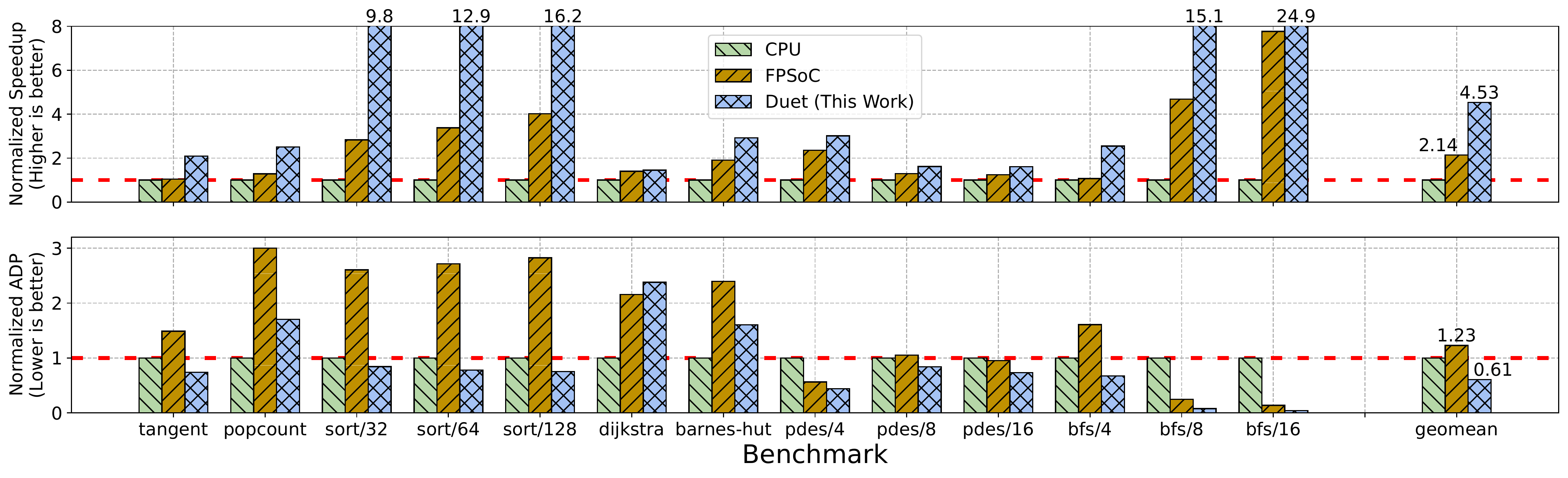}
    \setlength{\abovecaptionskip}{4pt}
    \caption{Normalized Speedup and ADP of Application Benchmarks}
    
    \begin{Center}
    \fontsize{9}{10}\selectfont
    \vspace{-0.5\baselineskip}
    \texttt{sort/N} indicates the array size of the sorting accelerator;
    \texttt{pdes/N} and \texttt{bfs/N} indicates the number of cores sharing the accelerator.
    \vspace{-1.2\baselineskip}
    \end{Center}
    
    \label{fig:speedup}
\end{figure*}

%% file: tables/app_util.tex
\begin{table}[t]

    \small
    \setlength\tabcolsep{6pt}
    \centering
        
        \setcounter{notenumber}{0}

    \newcommand{\mynote}{\textsuperscript{\fnsymbol{notenumber}}}
    \newcommand{\newnote}{\stepcounter{notenumber}\mynote{}}

        \caption{\centering Clock Frequency and Area of Soft Accelerators}
        
    \begin{threeparttable}[t]
        \begin{tabular}{|c|c|c|c|c|}
        
        \hline
        
        \multirow{2}{*}{Benchmark} &
        Max. Freq &
        Norm. &
        \multicolumn{2}{c|}{Resource Utilization} \\
        
        \cline{4-5}
        
        &
        (MHz) &
        Area~\newnote{} &
        CLB~\newnote{} &
        BRAM \\
        
        \hline \hline
        
        Ariane + &
        \multirow{2}{*}{1000} &
        \multirow{2}{*}{1.0} &
        \multirow{2}{*}{-} &
        \multirow{2}{*}{-} \\
        
        P-Mesh Socket &
        &
        &
        &
        \\
        
        \hline
        
        Tangent &
        282 &
        0.47 &
        0.84 &
        0 \\
        
        \hline
        
        Popcount &
        189 &
        2.77 &
        0.83 &
        0.56 \\
        
        \hline
        
        Sort (32) &
        228 &
        6.29 &
        0.30 &
        0.76 \\
        
        \hline
        
        Sort (64) &
        234 &
        8.10 &
        0.27 &
        0.92 \\
        
        \hline
        
        Sort (128) &
        228 &
        10.27 &
        0.27 &
        0.92 \\
        
        \hline
        
        Dijkstra &
        127 &
        1.94 &
        0.96 &
        0.31 \\
        
        \hline
        
        Barnes-Hut &
        85 &
        14.22 &
        0.99 &
        0.05 \\
        
        \hline
        
        BFS &
        208 &
        1.24 &
        0.61 &
        0.75 \\
        
        \hline
        
        %
        
        PDES &
        126 &
        2.77 &
        0.47 &
        0.56 \\
        
        \hline
        \end{tabular}
        
        \setcounter{notenumber}{0}
        
        \begin{tablenotes}
            \item [\stepcounter{notenumber}\fnsymbol{notenumber}] \footnotesize Total silicon area needed by the eFPGA to implement the accelerator, normalized to 1x Ariane + 1x P-Mesh Socket
            \item [\stepcounter{notenumber}\fnsymbol{notenumber}] \footnotesize \textit{Configurable Logic Block}, including LUTs and flip-flops
        \end{tablenotes}
        
        
    \end{threeparttable}
        \label{tab:eval-app}

\end{table}

%% file: tex/related.tex
\section{Related Work} \label{sec:related_work}

\subsection{Standalone FPGA-Based Accelerators}

FPGAs are being used in a fast-growing range of application domains due to their programmability and close-to-ASIC performance.
On a smaller scale, standalone FPGAs are used to accelerate database queries~\cite{fpga-database}, image processing~\cite{fpga-image-processing}, network filtering~\cite{fpga-network}, and most popularly, deep learning applications~\cite{fpga-dl, fpga-dl2, fpga-lstm}.
On the other end of the spectrum, FPGAs are massively deployed in the Cloud, either offered directly as an on-demand computing service such as the Amazon AWS F1 instances~\cite{amazon-f1}, or used to build cloud-scale accelerators such as Microsoft Catapult~\cite{ms-catapult} and BrainWave~\cite{ms-brainwave}.
In these systems, the FPGAs are often integrated as PCIe peripherals or separate machines on the datacenter network.
As explained in Sec.~\ref{sec:introduction}, such systems excel in accelerating at the application level, but are insufficient for fine-grained acceleration due to high CPU-FPGA communication overhead.

\subsection{Same-Package CPU-FPGA Hybrids and FPSoCs} \label{sec:related-fpsoc}

Addressing the CPU-FPGA communication bottleneck, academia and industry have explored tighter integration of the two.
Intel Harp~\cite{intel-harp} combines an Intel Xeon multi-core CPU and an Altera FPGA in one package, and bridges the two with QPI, a point-to-point interconnect with cache coherence support.
Field-Programmable System-on-Chips (FPSoC) achieves even tighter integration by integrating processors and FPGAs on the same chip, similar to Duet at a high level.
Many commodity FPSoCs~\cite{xilinx-zynq, altera-cyclone, smartfusion2, polarfire, quicklogic-eos-s3} and academic FPSoCs~\cite{epfl-arnold, fabulous, harvard-vlsi-19} support full or partial cache coherence.
For example, the Xilinx Zynq-7000 employs the AXI4 ACP interface~\cite{axi4} which supports uni-directional cache coherence (I/O coherency).
The Xilinx Zynq UltraScale+ MPSoC~\cite{xilinx-zynq-up} and Versal ACAP~\cite{xilinx-acap} provide full coherence support with the ACE protocol~\cite{axi4}.

There are two key differences between Duet and FPSoCs.
First, as shown in Fig.~{\ref{fig:arch-comp-fpsoc}}, the processor subsystem and the eFPGA in FPSoCs are separated by a centralized interconnect, and the hardware cache hierarchy resides in the processor subsystem.
Such organization is reasonable for an FPGA-centric, dual-core architecture, but cannot scale to a larger number of cores.
Second, in the absence of FPGA-side hardware caches, it becomes the accelerator designers' burden to design and verify soft caches that comply with the coherence protocol, for example ACE~{\cite{axi4}}.
Besides increasing design complexity, the soft caches' direct participation in cache coherence slows down the entire cache system because they run in the slow clock domain.
Moreover, since the soft caches have access to the micro-architecture states of the NoC, the system cannot restrain faulty or malicious behaviors of the soft caches.

\subsection{Near- and In-Processor Reconfigurable Computing} \label{sec:rfu}

Dating back to the early days of reconfigurable computing, computer architects have proposed to integrate Reconfigurable Fabrics (RF) very close to or even into processors.
Garp~\cite{garp} places an RF between a processor and its private cache, making the two compute units share the entire memory system.
Chimaera~\cite{chimeara} and PRISC~\cite{PRISC} embed Reconfigurable Functional Units (RFUs) directly into a processor's datapath, enabling post-fabrication customization of the Instruction Set Architecture (ISA).
The Post-Fabrication Microarchitecture~\cite{PFM} couples an RF with a superscalar core and allows the RF to observe and microarchitecturally intervene at key pipeline stages. 
These systems have their advantages but conflict with our \textit{plug-and-play integration} goal as they mandate the redesign of the processors.

\subsection{Coherence Protocols for Accelerators}

Hardware cache coherence has been widely adopted in multi-processor systems as they provide not only programmability but also performance.
In the context of hardware acceleration, recent studies have also shown benefits of employing hardware cache coherence and have proposed a variety of solutions.
FUSION~\cite{fusion} is a hierarchical, timestamp-based coherence protocol emphasizing data transfer between accelerators;
Crossing Guard~\cite{crossing-guard} proposes the use of a hard cache transducer to decouple the accelerator coherence protocol and the processor (host) protocol, as well as to incorporate fault-tolerance and memory translation;
Spandex~\cite{spandex} offers flexible write strategy and granularity, adapting to highly heterogeneous architectures.
While all of these proposals achieve hardware cache coherence, they are not optimized for FPGA-based accelerators in that they all require direct participation of the accelerators.
However, through the use of the Proxy Cache, Duet can be adapted to these coherence protocols, pretending to be a "fast" accelerator to the NoC and hiding the slow clock domain from the coherence protocol.

Accelerator coherence standards and interconnects are also emerging in industry, e.g., CCIX~\cite{ccix}, OpenCAPI~\cite{opencapi}, and CXL~\cite{cxl}.
These proposals target the board level or above and are really optimized for coarse-grained acceleration.


%% file: tex/conclusion.tex
\section{Conclusion}

In this work, we present Duet, a scalable, manycore-FPGA architecture with non-intrusive, coherently-integrated, embedded FPGAs that is optimized for \textit{fine-grained acceleration} and \textit{hardware augmentation} in the broad general-purpose domains.
By promoting the eFPGA to a first-class citizen on chip, Duet enables the eFPGA to access the NoC and to participate in bi-directional cache coherence just as any other processor. 
The novel, lightweight, Duet Adapters reduce critical communication overheads by placing the Proxy Caches and Shadow Registers in the faster processor clock domain.
Our evaluation shows that Dolly, an RTL-level implementation of Duet, can reduce the latency of processor-accelerator communications by up to 82\% and increase the bandwidth by up to 9.5x, stably across a wide range of eFPGA clock frequencies.
Selected benchmarks leveraging Duet-enabled soft accelerators
show up to 24.9x speedup over processor-only baselines and up to 4x over FPSoCs.
Dolly and its toolchain have been open-sourced and available at \url{https://github.com/PrincetonUniversity/Duet}.